\documentclass[12pt]{article}
\usepackage{amssymb}

\usepackage{amsmath,amsfonts,xypic}
\usepackage{amscd}
\usepackage[all]{xy}
\usepackage{graphicx}

\advance \textheight by \topskip
\makeatletter
\@addtoreset{equation}{section}

\makeatother

\newcommand{\beq}{\begin{equation}}
\newcommand{\eeq}{\end{equation}}
\newcommand{\beqa}{\begin{eqnarray}}
\newcommand{\eeqa}{\end{eqnarray}}
\newcommand{\noi}{\noindent}

\def\>{\rangle}
\def\<{\langle}

\begin{document}

\title{
{\bf  On the ternary complex analysis and its applications }  }

\bigskip \bigskip 

\vspace{0.7in} 
\begin{titlepage}
\author{ 
{\sf   L. N. Lipatov} \thanks{e-mail:
lipatov@mail.desy.de}$\,\,$${}^{a}$,
{\sf  M. Rausch de Traubenberg }\thanks{e-mail:
rausch@lpt1.u-strasbg.fr}$\,\,$${}^{b}$
and
{\sf G. G. Volkov}\thanks{e-mail:
Guennadi.Volkov@cern.ch}$\,\,$$^{a,b,c}$
\\
{\small ${}^{a}${\it PNPI}} \\
{\small {\it St-Petersburg, Russia}}  \\ 
{\small ${}^{b}${\it
Laboratoire de Physique Th\'eorique, CNRS UMR  7085,
Universit\'e Louis Pasteur}}\\
{\small {\it  3 rue de
l'Universit\'e, 67084 Strasbourg Cedex, France}}\\
{\small ${}^c$ {\it LAPTH, Universit\'e de Savoie }}\\
{\small {\it 9 chemin de Bellevue, B.P. 110, F74941 Annecy-le-Vieux, Cedex, France}}
}
\date{\today}
\maketitle
\vskip-1.5cm

\maketitle

\vspace{.7in}

\begin{abstract}
Previouly a possible extension of the complex number, together
with its connected trigonometry was introduced. In this paper
we focuss on the simplest case of ternary complex numbers. 
Then, some types of holomorphicity adapted 
to the ternary complex numbers  and the corresponding results
upon integration  of differential forms are given. Several physical 
applications are given, and in particuler one type of holomorphic 
function gives rise to a new form of stationary magnetic field.
The movement of a monopole  type object  
in this field  
is then studied and shown to be integrable.
The monopole scattering in the ternary field is finally studied.

\end{abstract}

\end{titlepage}

\section{ Geometric origin of complex numbers.}
The complex number theory is a seminal field in mathematics having many
applications to geometry, group theory, algebra and also to the classical
and quantum physics. Geometrically, it is based on the complexification of
the $\mathbb {R}^{2}$ plane. The existence of similar structures in higher
dimensional spaces is interesting for phenomenological applications.

As a first step toward an extension of the complex numbers, in a 
series of papers mathematicians studied a three dimensional space with
linear arc defined by $d^3 s = d^3 x  + d^3 y + d^3 z 
-3 d  x  d y  d  z$. This study was initiated by two papers of
P. Appell \cite{appell}, written in 1877,
 where he  introduced some generalisation of the usual trigonometric
functions which are closely related to the linear arc defined above.
Then, the geometrical properties
of these three-dimensional spaces was firstly  studied
by P. Humbert in a series of small papers
(mostly in Comptes Rendus Acad\'emie des Sciences, Paris)
and then by J. Devisme. 
One can see for instance \cite{humbert1, humbert2, humbert3, devisme1,
devisme2} and references theirin.
In 1933 and  1940 J. Devisme \cite{devisme1,devisme2}
wrote  very interesting
reviews on the subjet 
summarizing all the results obtained at that time (especially in the
first one).
However, all the results of  these papers were  obtained 
without any reference to some extension of the complex
number (except in a small remark in the first  paper of Appell \cite{appell}).
 
It turns out, that there is a natural way to construct a generalized version
of complex numbers in $\mathbb {R}^{n}$ spaces ($n=3,4,...)$. 
Such numbers
were   introduced in 
 \cite{mc1, mc2}. In these papers, the authors suggested to use
group theoretical methods to define these numbers, and in particular, 
the cyclic group $C_n$ was used for the ``complexification'' of
 $\mathbb R^n$. The simplest case beyond the complex numbers,
corresponding to $n=3$ is  called ternary complex numbers, 
$\mathbb T_n \mathbb C$ is the algebra generated by one canonical generator
 $q$ satisfying $q^3=1$.  The consideration
of these numbers  allows to
rediscover in a more efficient way most of the
results given in \cite{devisme1}. These numbers gave also  a new insight
on the achievements summarized in \cite{devisme1} together with 
some new relations.
Very foundamental results obtained within the complex
number approach (Virasoro and  Kac-Moody algebras in (super)string 
theory and
integrable models, {\it etc.}) give strong motivation to study with another
point of view and more details
the ternary complex numbers as well as their symmetric
and geometric properties.
The ternary complex numbers might be related to some new symmetries  
in high energy physics  allowing to explain the three color- 
\cite{kerner} and three family-\cite{volkov}
problems of the Standard Model.

In this paper we are going to investigate new aspects of the ternary complex
analysis based on the ``complexification'' of $\mathbb {R}^{3}$ 
space \cite{mc1,mc2}. 
The use of the cyclic $\mathit{C}_{3}$ group for this purpose is a natural
generalization of the similar application of the $\mathit{C}_{2}={Z}_{2}$
group in two dimensions. It is known that the complexification of 
$\mathbb {R}^{2}$
allows to introduce the new geometrical objects - the Riemannian surfaces.
The Riemannian surfaces are defined as a pair $(M,\Omega)$, where $M$ is a
connected two-dimensional manifold and $\Omega$ is a complex structure on $M$.
Well-known examples of Riemann surfaces are the complex plane- 
$\mathbb {C}$,
Riemann sphere - ${CP}^{1}=\mathbb C\cup \left\{{\infty }\right\}$, complex tori- 
$ {T}=\mathbb {C}/{\Gamma 
}$, $\Gamma :=n\lambda _{1}+n\lambda _{2}:n,m\in {Z}$, $\lambda _{1,2}\in {C}
$, {\it etc.}

For the ternary complexification of the vector space, 
$\mathbb {R}^{3}$, one uses
its cyclic symmetry subgroup $\mathit{C}_{3}={R}_{3}$ \cite{mc1, mc2}. In
the physical context the elements of this subgroup  are actually
spatial rotations through a restricted set of angles, $0,2\pi /3,4\pi /3$
around, for example, the $x_{0}$-axis. After such rotations the coordinates, 
$x_{0},x_{1},x_{2}$, of the point in $\mathbb {R}^{3}$ are linearly related with the
new coordinates, $x_{0}^{\prime },x_{1}^{\prime },x_{2}^{\prime }$ which can
be realized by the $3\times 3$ matrices. 
The vector representation $D^{V}$ is defined through
the following three orthogonal matrices: 
\[
R^{V}(1)=O(0)=\left( 
\begin{array}{ccc}
1 & 0 & 0 \\ 
0 & 1 & 0 \\ 
0 & 0 & 1
\end{array}
\right) \,, 
\]

\[
R^{V}(q)=O(2\pi /3)=\left( 
\begin{array}{ccc}
1 & 0 & 0 \\ 
0 & -1/2 & \sqrt{3}/2 \\ 
0 & -\sqrt{3}/2 & -1/2
\end{array}
\right) \,, 
\]

\[
R^{V}(q^{2})=O(4\pi /3)=\left( 
\begin{array}{ccc}
1 & 0 & 0 \\ 
0 & -1/2 & -\sqrt{3}/2 \\ 
0 & \sqrt{3}/2 & -1/2
\end{array}
\right) \,. 
\]

\noi
These matrices realize the group multiplication rules
due to the relations $%
R^{V}(q^{2})=(R^{V}(q))^{2}$ and $(R^{V}(q))^{3}=R^{V}(1)$. The
representation is faithful because the kernel of its homomorphism consists
only of identity: $\text{Ker} R=1$.

Let us introduce the matrix

\begin{eqnarray} \label{caract}
\hat x = \sum \limits_{i=0}^3 x_i  R^V(q^i)= \left ( 
\begin{array}{ccc}
x_0+x_1+x_2 & 0 & 0 \\ 
0 & x_0-1/2(x_1+x_2) & -\sqrt{3}/2(x_1-x_2) \\ 
0 & \sqrt{3}/2(x_1-x_2 ) & x_0-1/2(x_1+x_2)
\end{array}
\right ).
\end{eqnarray}

\noi
The determinant of this matrix is

\begin{eqnarray}
\text{det} (\hat x)=x_0^3+x_1^3+x_2^3-3x_0x_1x_2\,,
\end{eqnarray}

\noi
which is precisely the linear arc defined in \cite{humbert1,humbert2,humbert3,
devisme1,devisme2}.

The cyclic group $\mathit{C}_{3}$ has three conjugation classes, $1$, $q$
and $q^{2}$, and, respectively, three one dimensional irreducible
representations, ${R}^{(i)}$, $i=1,2,3$. We write down the table of their
characters \cite{jones}

\[
\begin{array}{c|ccc}
\mathit{C}_{3} &1 & q & q^{2}   \\ \hline
R^{(1)} & 1 & 1 & 1  \\ 
R^{(2)} & 1 & j & j^{2}  \\ 
R^{(3)} & 1 & j^{2} & j 
\end{array}
\,\,\,,\,\,\,\,j^{3}=1\,, 
\]
where $R^{(1)}$ is the trivial representation, whereby each elements is
mapped onto unit, \textit{i.e.} for $R^{(1)}$ the kernel is the whole group, 
$\mathit{C}_{3}$. For $R^{(2)}$ and $R^{(3)}$ the kernels can be identified
with unit element, which means that they are faithful representations,
isomorphic to $\mathit{C}_{3}$.

We remind that for the cyclic group $\mathit{C}_{2}$ there are two
conjugation classes, $1$ and $i$ and two one-dimensional irreducible
representations:

\[
\begin{array}{c|cc|c}
\mathit{C}_{2} & 1 & i &  \\ \hline
R^{(1)} & 1 & 1 & z \\ 
R^{(2)} & 1 & -1 & \bar{z}
\end{array}
\,\,. 
\]

Based on the character table and on \eqref{caract} one can obtain 
\[
\xi ^{V}=\left(\xi ^{V}(1)=\text{Tr}(R^V(1)),\xi^{V}(q)=\text{Tr}(R^V(q)),
\xi ^{V}(q^{2})=\text{Tr}(R^V(q^2))\right)=(3,0,0)\,, 
\]
which demonstrates how the vector representation $R^{V}$ decomposes in the
irreducible representations $R^{(i)}$: 
\[
\xi ^{V}=\xi ^{(1)}+\xi ^{(2)}+\xi ^{(3)} 
\]
or 
\[
R^{V}=R^{(1)}\oplus R^{(2)}\oplus R^{(3)}. 
\]
The combinations of coordinates on which $R^{V}$ acts irreducible are given
below

\[
\left( 
\begin{array}{c}
z \\ 
\tilde{z} \\ 
\tilde{\tilde{z}}
\end{array}
\right) =\left( 
\begin{array}{ccc}
1 & 1 & 1 \\ 
1 & j & j^{2} \\ 
1 & j^{2} & j
\end{array}
\right) \left( 
\begin{array}{c}
x_{0} \\ 
x_{1}q \\ 
x_{2}q^{2}
\end{array}
\right) \,. 
\]

The content of the paper is as follows. In the next section we remind the
basic results on ternary complex numbers obtained in \cite{mc1,mc2}
useful in the sequel, in particular we introduce some functions,
which extend the usual sine and cosine functions,  having generalized
analytical properties. These functions coincide with those introduced
by Appell \cite{appell}.
 Section 3 is devoted to the formulation of some types of
holomorphicity adapted to the ternary complex numbers and to
the integration of differential forms. In particular
a generalization of the logarithm   is  introduced, and it 
turns
out that corresponding functions are directly related to
expressions extending the usual sine and cosine.
In this section we give also a physical interpretation
of one ternary differential form. It appears that this form
can be interpreted as a magnetic field generated by a monopole
string.
In section 4 we construct a ``ternary'' Newton mechanics and
study the movement of a monopole  type object  
in the magnetic field constructed in the 
previous sub-section. It turns out, that the Newton equation is integrable,
which allows us to investigate the monopole scattering in the ternary field.

\section{Ternary complex numbers}

In Refs. \cite{mc1},\cite{mc2} an $n-$dimensional commutative extension of
complex numbers was introduced together with functions having generalized
analytical properties (see also \cite{appell,devisme1}). 
The algebra of the multicomplex numbers is defined
over the field of real numbers; it is generated by one element $q$
satisfying the condition $q^{n}=-1$. Below we remind some results obtained
in \cite{mc1} (see also \cite{devisme1})
for the simplest non-trivial case $n=3$. The class of these
ternary complex numbers is denoted by ${\mathbb T}_{3}\mathbb C$. 
Since $3$ is an odd number
the generator $q$ can be equivalently normalized in such a way, that 
\[
q^{3}=1\,, 
\]
which leads to some minor modifications in formul\ae \ derived in Ref. \cite
{mc1}.

A ternary complex number $z\in \mathbb {T}_{3}\mathbb C$ 
is expressed in terms of $%
x_{0},x_{1},x_{1}\in \mathbb R$ as follows 
\[
z=x_{0}+x_{1}q+x_{2}q^{2}  \, .
\]
Formally ${\mathbb T}_{3}\mathbb C$ can be defined by

\[
{\mathbb T}_{3}\mathbb C=\mathbb R[X]/(1-X^{3})\,. 
\]

\noindent Since $1-X^{3}=(1-X)(1+X+X^{2})$ we have some interesting
algebraic properties for ${\mathbb T}_{3}\mathbb C$:

\begin{itemize}
\item[1]  . ${\mathbb T}_{3}\mathbb C$ is not simple : 
${\mathbb T}_{3}\mathbb C=I_{1}\oplus I_{2}$ with two
ideals $I_{1},I_{2}$;

\item[2]  . $\forall z_{1}\in I_{1},z_{2}\in I_{2}$ $z_{1}z_{2}=0$;

\item[3]  . $I_{1}\cong \mathbb R \lbrack X]/(1-X)\cong \mathbb R,
\,I_{2}\cong \mathbb R\lbrack
X]/(1+X+X^{2})\cong \mathbb C\,$.
\end{itemize}

\noi
This can be proven easily. Indeed, let us introduce

\beqa
\label{EKI}
K_0&=&\frac{1}{3}\left( 1+q+q^{2}\right)\,, \nonumber \\
I&=&\frac{%
1}{\sqrt{3}}\Big((1-q)-(1-q)^{2}\Big)=\frac{1}{\sqrt{3}}(q-q^{2})\,, \\
E_0&=&-I^{2}=\frac{1}{3}(2-q-q^{2}) \nonumber
\eeqa
and denote by $\rho $ the projection from $%
\mathbb R[X]\rightarrow {\mathbb T}_{3}\mathbb C$ and by 
$q=\rho (X)$ the image of $X$ in ${\mathbb T}_{3}\mathbb C$. 
Since by definition $\mbox{Ker}\rho =\left\langle 1-X^{3}\right\rangle $
it is clear that $(1-q)(1+q+q^{2})=0.$ If we define 
$I_{1}\subseteq {\mathbb T}_{3}\mathbb C$
to be the one-dimensional sub-algebra generated by $K_{0}
,\ $ and $I_{2}\subseteq {\mathbb T}_{3}\mathbb C$ 
to be the two-dimensional sub-algebra generated by $I$,  it is easyly to
verify the relations

\begin{eqnarray}  \label{IEK}
K_0^n=K_0, E_0^n=E_0, n \ge 1, \ I^2=-E_0, K_0 E_0=0, E_0 I = I.
\end{eqnarray}

\noindent This proves the above statements 1-2-3. Note also that we have the
relation $e^{\frac{2\pi }{3}I}=q.$

A ternary complex number can be presented in two different forms

\begin{eqnarray}  \label{basis}
z=x_0 + x_1 q + x_2 q^2= (x_0+x_1+x_2) K_0 +(x_0 -\frac12 (x_1 - x_2)) E_0 + 
\frac{\sqrt{3}}{2}(x_1-x_2) I,
\end{eqnarray}

\noindent with $x=(x_0+x_1+x_2) K_0 \in I_1, w=(x_0 -\frac12 (x_1 - x_2))
E_0 + \frac{\sqrt{3}}{2}(x_1-x_2) I \in I_2.$

Furthermore, since ${\mathbb T}_{3}\mathbb C=I_{1}\oplus I_{2}$ 
with $I_{1}\ncong I_{2}$ any
automorphism of ${\mathbb T}_{3}\mathbb C$ 
preserves the two ideals $I_{1}$ and $I_{2}$. But 
$I_{1}\cong \mathbb R$ and $I_{2}\cong \mathbb C$, 
which means that there is only one
automophism of ${\mathbb T}_{3}\mathbb C$ given by 
$q\leftrightarrow q^{2}$ (or $%
I\rightarrow -I,E_{0}\rightarrow E_{0}$ and $K_{0}\rightarrow K_{0}$ \textit{%
i.e.} for $w\in I_{2}$ we have $w=xE_{0}+yI\rightarrow xE_{0}-yI$). This
automorphism enables us to define an adapted ``modulus'' on 
$z\in {\mathbb T}_{3}\mathbb C$.
Namely, for any $z\in {\mathbb T}_{3}\mathbb C$ such that 
$z=x+w,$ with $x\in I_{1},w\in
I_{2}$ we set

\begin{eqnarray}  \label{modulus}
\parallel z \parallel^3 = \parallel x \parallel_{_{I_1}} \parallel w
\parallel_{_{I_2}}^2 &=&(x_0 + x_1 + x_2) \Big((x_0 -\frac12 (x_1 - x_2))^2+ 
\frac{{3}}{2}(x_1-x_2)^2\Big) \\
&=&(x_0 +x_1+x_2)(x_0 +jx_1+j^2x_2)(x_0 +j^2x_1+jx_2)  \nonumber \\
&=&x_0^3 + x_1^3 + x_2^3 -3 x_0 x_1 x_2 \nonumber
\end{eqnarray}

\noindent with $j=e^{\frac{2i\pi }{3}}$. It is interesting that a different
definition of $\parallel z\parallel $ can be given. Indeed, let us consider
three isomorphic copies of  ternary complex numbers

\begin{eqnarray}
{\mathbb T}_{3}\mathbb C &=&
\left\{ z=x_{0}+x_{1}q+x_{2}q^{2},\ x_{0},x_{1},x_{2}\in
\mathbb R\right\} \,,  \nonumber \\
\tilde{\mathbb T}_{3}\mathbb C &=&
\left\{ \tilde{z}=x_{0}+x_{1}jq+x_{2}j^{2}q^{2},\
x_{0},x_{1},x_{2}\in \mathbb R\right\} \,, \\
\tilde{\tilde{\mathbb T}}_{3}\mathbb C &=&\left\{ {\tilde{\tilde{z}}}%
=x_{0}+x_{1}j^{2}q+x_{2}jq^{2},\ x_{0},x_{1},x_{2}\in \mathbb R\right\} \,. 
\nonumber
\end{eqnarray}

\noindent The ternary isomorphism \ $\tilde{}$ \ is the mapping

\[
{\mathbb T}_{3}\mathbb C\rightarrow 
\tilde{\mathbb T}_{3}\mathbb C\rightarrow 
\tilde{\tilde{\mathbb T}}%
_{3}\mathbb C\rightarrow {\mathbb T}_{3}\mathbb C\,. 
\]

\noindent For further use, note that for elements $z,\tilde{z}$ and $\tilde{%
\tilde{z}}$ of the algebras\footnote{%
If we complexify the ternary complex numbers 
${\mathbb T}_{3}\mathbb C^{\mathbb C}={\mathbb T}_{3}
\mathbb C\otimes_{\mathbb R}\mathbb C$, 
the three above copies become identical and $\ \ \tilde{}\ $ is an
automorphism.} ${\mathbb T}_{3}\mathbb C,
\tilde{\mathbb T}_{3}\mathbb C$ and $\tilde{\tilde{\mathbb T}}_{3}\mathbb C$ we
have $\tilde{z}+\tilde{\tilde{z}}=2 x_{0}-x_{1}q-x_{2}q^{2}
\in {\mathbb T}_{3}\mathbb C$, $%
\tilde{z}{\tilde{\tilde{z}}}%
=(x_{0}^{2}-x_{1}x_{2})+(x_{2}^{2}-x_{0}x_{1})q+(x_{1}^{2}-x_{2}x_{0})q^{2}%
\in {\mathbb T}_{3}\mathbb C$. We also have 

\begin{eqnarray}  \label{decomposition}
x_0=\frac13\left(z+\tilde z + \tilde {\tilde {z}}\right),\ x_1=\frac{q^2}{3}%
\left(z+j^2\tilde z + j\tilde {\tilde {z}}\right), \ x_2=\frac{q}{3}%
\left(z+j\tilde z + j^2\tilde {\tilde {z}}\right).
\end{eqnarray}
The ternary mapping allows to give an alternative definition of the
pseudo-norm on ${\mathbb T}_3\mathbb C $ by

\[
\begin{array}{lcll}
\parallel \ \parallel : & {\mathbb T}_{3}\mathbb C
\otimes \tilde{\mathbb T}_{3}\mathbb C\otimes \tilde{%
\tilde{\mathbb T}}_{3}\mathbb C & \rightarrow & \mathbb R\,, \\ 
& z\otimes \tilde{z}\otimes {\tilde{\tilde{z}}} & \mapsto & \parallel
z\parallel ^{3}=z\tilde{z}{\tilde{\tilde{z}}}%
=x_{0}^{3}+x_{1}^{3}+x_{2}^{3}-3x_{0}x_{1}x_{2}
\end{array}
\]

\noindent which coincide with \eqref{modulus}. Thus, $\parallel z\parallel
=0 $  \textit{if and only if} 
$z$ belongs to $I_{1}$ or to $I_{2}$. A ternary
complex number is called non-singular if $\parallel z\parallel \neq 0$.
From now on we also denote $|z|$ the modulus of $z$. The modulus
introduced here coincide with the one considered in \cite{humbert1,humbert2,
humbert3,devisme1,devisme2}.

It was proven in \cite{mc1} that any non-singular ternary complex number $%
z\in {\mathbb T}_{3}\mathbb C$ can be written in the ``polar form'':

\begin{eqnarray}  \label{polar}
z= \rho e^{\varphi_1 q + \varphi_2 q^2}=\rho e^{\theta(q-q^2) +
\varphi(q+q^2)}
\end{eqnarray}

\noindent with $\rho = |z| =
\sqrt[3]{x_{0}^{3}+x_{1}^{3}+x_{2}^{3}-3x_{0}x_{1}x_{2}%
}\in \mathbb  R,\theta \in \lbrack 0,2\pi/\sqrt{3} \lbrack ,
\varphi \in \mathbb R $. The
combinations $q-q^{2}$ and $q+q^{2}$ generate in the ternary space
respectively compact and non-compact directions. Using \eqref{IEK} and $%
q+q^{2}=2K_{0}+E_{0}$, we can rewrite \eqref{polar} in the form

\begin{eqnarray}  \label{mus}
z=\rho [m_0(\varphi_1,\varphi_2) + m_1(\varphi_1,\varphi_2)q +
m_2(\varphi_1,\varphi_2)q^2]
\end{eqnarray}

\noindent where the multi-sine functions \cite{mc1, int} are given below
(in \cite{appell, devisme1} these functions are denoted $P,Q,R$).

\begin{eqnarray}
\label{mus2}
m_{k}(\varphi _{1},\varphi _{2}) &=&\frac{1}{3}\left( e^{\varphi
_{1}+\varphi _{2}}+2e^{-\frac{1}{2}\left( \varphi _{1}+\varphi _{2}\right)
}\cos \big(\frac{\sqrt{3}}{2}(\varphi _{1}-\varphi _{2})-\frac{2k\pi }{3}%
\big)\right)  \label{mus-fct} \\
&=&\frac{1}{3}\left( e^{\varphi _{1}+\varphi _{2}}+j^{2k}e^{j\varphi
_{1}+j^{2}\varphi _{2}}+j^{k}e^{j^{2}\varphi _{1}+j\varphi _{2}}\right) ,\ \
\ k=0,1,2\,\,.  \nonumber
\end{eqnarray}

\noindent They satisfy the relations

\begin{eqnarray}
\frac{\partial }{\partial \varphi _{\ell }}m_{k}(\varphi _{1},\varphi _{2})
&=&m_{k-\ell }(\varphi _{1},\varphi _{2}) \,, \\
m_{k}(\varphi _{1}+\psi _{1},\varphi _{2}+\psi _{2}) &=&m_{0}(\varphi
_{1},\varphi _{2})m_{k}(\psi _{1},\psi _{2})+m_{1}(\varphi _{1},\varphi
_{2})m_{k-1}(\psi _{1},\psi _{2})  \nonumber \\
&+&m_{2}(\varphi _{1},\varphi _{2})m_{k-2}(\psi _{1},\psi _{2})\,,  \nonumber
\end{eqnarray}

\noindent where the indices are defined modulo $3$ and

\begin{eqnarray}
m_0^3(\varphi_1,\varphi_2)+m_1^3(\varphi_1,\varphi_2)+m_2^3(\varphi_1,%
\varphi_2)-
3m_0(\varphi_1,\varphi_2)m_1(\varphi_1,\varphi_2)m_2(\varphi_1,\varphi_2)
=1\,.
\end{eqnarray}

Since for the product of two ternary complex numbers we have $\parallel
zw\parallel =\parallel z\parallel \parallel w\parallel $ the set of
unimodular ternary complex numbers preserves the cubic form \cite{mc1, int}.
The continuous group of symmetry of the cubic surface $%
x_{0}^{3}+x_{1}^{3}+x_{2}^{3}-3x_{0}x_{1}x_{2}=\rho ^{3}$ is isomorphic to $%
SO(2)\times SO(1,1)$. We denote the set of unimodular ternary complex
numbers or the ``ternary unit sphere'' as $\mathbb TU(1)=
\left\{ e^{(\theta +\varphi
)q+(\varphi -\theta )q^{2}},\ 0\leq \theta <2\pi/\sqrt{3} ,\varphi \in
\mathbb R\right\} \sim \mathbb T S^{1}$. This surface is also called
the Appell sphere \cite{devisme1}.

\begin{center}
\includegraphics[scale=0.55]{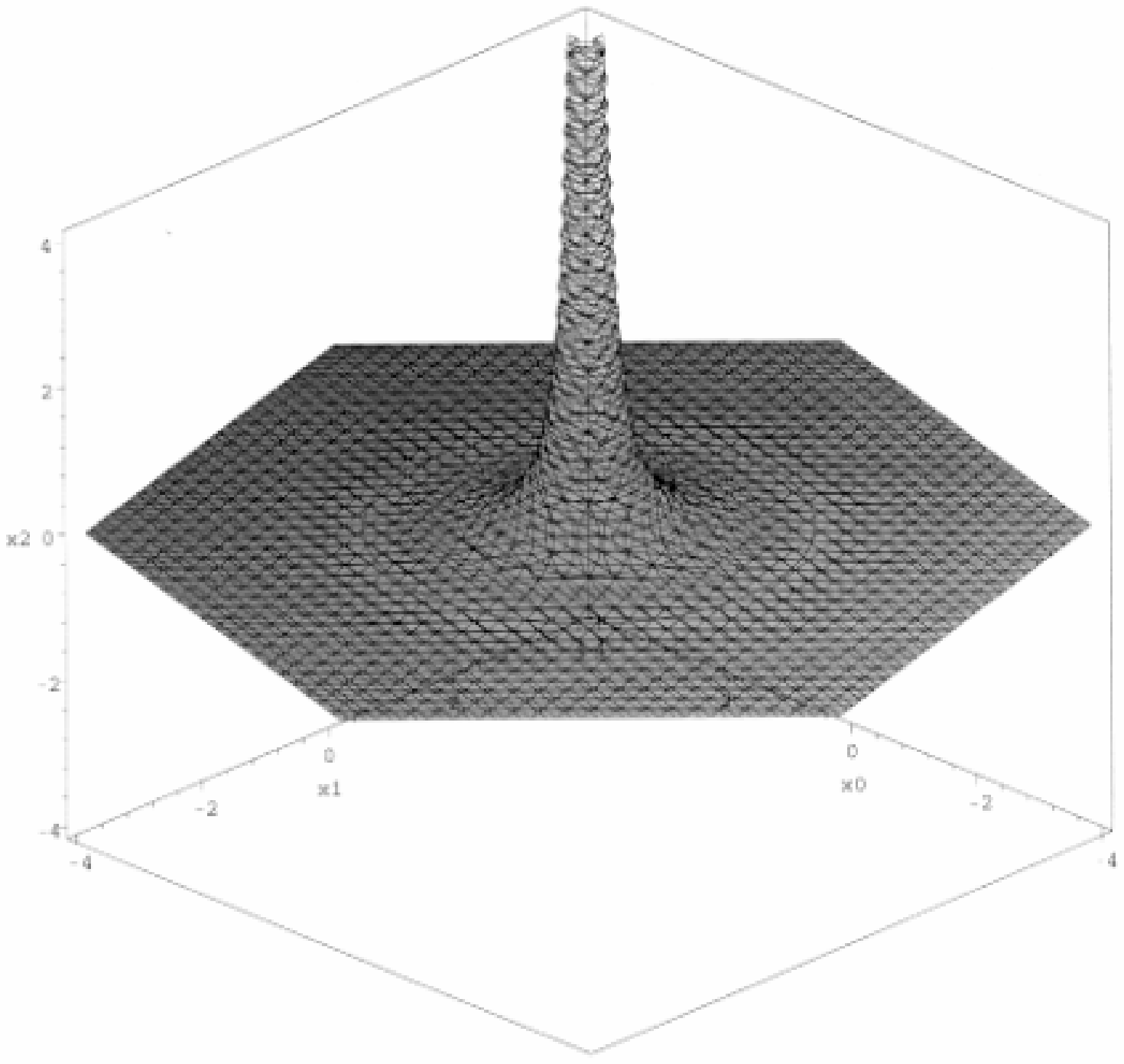}

{\small \textbf{Fig 1}: Cubic surface $x_0^3+x_1^3+x_2^3-3x_0x_1x_2=1$}
\end{center}

\noindent

Looking at
the above figure one can see, that the surface approaches
asymptotically the plane $x_{0}+x_{1}+x_{2}=0$ and the line $%
x_{0}=x_{1}=x_{2}$ is orthogonal to it. In $\mathbb T_{3}\mathbb C$ 
they correspond to the
ideals $I_{2}$ and $I_{1}$, respectively. The latter line will be called the
``trisectrice''.

One can construct the inner metric of this surface for the general case $\rho
\ne 0$. Let us introduce the new coordinate
$a=x_0+x_1+x_2$ and parametrise a point on the
circle around the trisectrice by its polar coordinates
$(r,\theta)$.
The surface in
these coordinates has the simple equation  $ar^{2}=\rho ^{3}$. It can be shown
that a point $M(x_0,x_1,x_2)$
 on the cubic surface can be  parametrised as follows

\beqa
\label{trumpet}
x_0=\frac13(\frac{\rho^3}{r^2}-2 r \cos \theta), \; 
x_1=\frac13(\frac{\rho^3}{r^2}+ r \cos \theta + 
\sqrt{3}r \sin \theta),\; 
x_2=\frac13(\frac{\rho^3}{r^2}+  r \cos \theta -
\sqrt{3}r \sin \theta).\nonumber \\
\eeqa
This gives 

\[
ds^{2}=\frac13(2+\frac{4\rho ^{6}}{r^{6}})dr^{2}+\frac23 r^{2}d\theta ^{2} 
\]
and corresponds to the Gauss curvature

\[
K=-\frac{12a^{4}}{(4a^{3}+\rho ^{3})^{2}}\,. 
\]
The Gauss
curvature is negative and tends to zero, when $a\rightarrow 0$ ($%
r\rightarrow \infty $). In this limit the metric becomes flat:

\begin{eqnarray}
ds^2=\frac23(d r^2+r^2 d\theta^2).
\end{eqnarray}

The $SO(2)\times SO(1,1)$ group of transformations is generated by the
multi-sine functions. In particular, in the special case where $\varphi =0$
the transformation in the compact direction is a rotation to the angle $%
\sqrt{3}\theta $ and for $\theta =0$ we have the dilatation in the
non-compact direction

\[
\begin{array}{l}
\varphi =0:\left\{ 
\begin{array}{l}
x_{0}+x_{1}+x_{2}\rightarrow x_{0}+x_{1}+x_{2} \\ 
x_{0}+jx_{1}+j^{2}x_{2}\rightarrow e^{i\sqrt{3}\theta
}(x_{0}+jx_{1}+j^{2}x_{2})
\end{array}
\right. \,, \\ 
\\ 
\theta =0:\left\{ 
\begin{array}{l}
x_{0}+x_{1}+x_{2}\rightarrow e^{2\varphi }(x_{0}+x_{1}+x_{2}) \\ 
x_{0}+jx_{1}+j^{2}x_{2}\rightarrow e^{-\varphi }(x_{0}+jx_{1}+j^{2}x_{2})
\end{array}
\right. \,.
\end{array}
\]

Let us consider now the following 
discrete transformation preserving the modulus $%
\parallel z\parallel $ of non-singular ternary complex numbers:

\begin{eqnarray}  \label{duality}
z= \rho e^{\varphi_1 q + \varphi_2 q^2} \to \bar z = \frac{\tilde z \tilde
{\tilde z }}{\parallel z \parallel} =\rho e^{-\varphi_1 q - \varphi_2 q^2}.
\end{eqnarray}

\noindent Obviously we have $\bar{\bar{z}}=z$. For multi-sine functions $%
m_{k}(\varphi _{1},\varphi _{2})$ the  symmetry \eqref{duality} leads to the
relations

\begin{eqnarray}
m_0(-\varphi_1,-\varphi_2)&=& m_0^2(\varphi_1,\varphi_2)-
m_1(\varphi_1,\varphi_2)m_2(\varphi_1,\varphi_2),  \nonumber \\
m_1(-\varphi_1,-\varphi_2)&=& m_2^2(\varphi_1,\varphi_2)-
m_0(\varphi_1,\varphi_2)m_1(\varphi_1,\varphi_2),  \nonumber \\
m_2(-\varphi_1,-\varphi_2)&=& m_1^2(\varphi_1,\varphi_2)-
m_0(\varphi_1,\varphi_2)m_2(\varphi_1,\varphi_2).  \nonumber
\end{eqnarray}

\noi
These relations are also given in \cite{devisme1}.

\section{Differential and holomorphic forms}
\subsection{Holomorpicity}
In this section we introduce notions of holomorphicity adapted to the
ternary complex numbers, some partial results was previously
given in \cite{mc2}. 
For a  point $a$ of a Riemann manifold the elements $dx_{0},dx_{1},dx_{2}$
form a basis of the cotangent space $T_{a}^{(1)}$. It should be relevant
to consider also the alternative basis given by  $(dz,d\tilde{z}%
,d\tilde{\tilde{z}})$. Thus, if one defines  a differentiable
function 

\beqa
\label{F}
F(x_{0},x_{1},x_{2})&=&f_0(x_0,x_1,x_2) + f_1(x_0,x_1,x_2)\ q + 
f_2(x_0,x_1,x_2)\  q^2 \\
&\equiv&g_0(z,\tilde z,\tilde {\tilde {z}}) +
g_1(z,\tilde z,\tilde {\tilde {z}})\  q +
g_2(z,\tilde z,\tilde {\tilde {z}})\  q^2\;,\nonumber
\eeqa

\noindent
it is possible to  construct the 1-form  
\begin{eqnarray}
dF =\frac{\partial F}{\partial x_{0}}dx_{0}+\frac{\partial F}{\partial
x_{1}}dx_{1}+\frac{\partial F}{\partial x_{2}}dx_{2}  
=\frac{\partial F}{\partial z}dz+\frac{\partial F}{\partial {\tilde{z}}}d{%
\tilde{z}}+\frac{\partial F}{\partial \tilde{\tilde{z}}}d\tilde{\tilde{z}}\,.
\end{eqnarray}
Here  we introduced

\beqa
\frac{\partial}{\partial z}&=&\frac13
\left( \frac{\partial}{\partial x_0} + 
 q^2\frac{\partial}{\partial x_1} +  \frac{\partial}{%
\partial x_2}\right)\,, \nonumber \\
\frac{\partial}{\partial \tilde z}&=&\frac13
\left( \frac{\partial}{\partial x_0} + 
j^{2} q^2\frac{\partial}{\partial x_1} + jq \frac{\partial}{%
\partial x_2}\right)\,, \\
\frac{\partial}{\partial {\tilde {\tilde z}}}&=&\frac13
\left( \frac{\partial}{\partial x_0} + 
j q^2\frac{\partial}{\partial x_1} + j^2 q \frac{\partial}{%
\partial x_2}\right). \nonumber 
\eeqa

In a full 
analogy with the usual complex analysis we are now able to
define holomorphic functions. Thus,
 $T_{a}^{(1)}$ is considered
as  a direct sum of three subspaces spanned 
by $dz$, 
$d{\tilde{z}}$ and $d\tilde{\tilde{z}}$: 
\[
T_{a}^{(1)}=T_{a}^{(1,0,0)}\oplus T_{a}^{(0,1,0)}\oplus T_{a}^{(0,0,1)}\,,
\]
or 
\[
T_{a}^{(1)}=T_{a}^{(1,0,0)}\oplus T_{a}^{(0,1,1)}. 
\]
We introduce now the following notations for one-forms
\begin{eqnarray}
\omega ^{(1,0,0)} &=&g_0(z,\tilde z ,{\tilde {\tilde z }}) \ dz\,, \ \
\omega ^{(0,1,1)} =g_1(z,\tilde z ,{\tilde {\tilde z }})
\ d\tilde{z}+g_2(z,\tilde z ,{\tilde {\tilde z }})
 \ d\tilde{\tilde{z}}\,,   \nonumber \\ 
\omega ^{(1)} &=&g_0(z,\tilde z ,{\tilde {\tilde z }})
 \ dz+ g_1(z,\tilde z ,{\tilde {\tilde z }})
 \ d\tilde{z}+g_2(z,\tilde z ,{\tilde {\tilde z }}) \ d\tilde{\tilde{z}}\,. 
\end{eqnarray}

\subsubsection{First type of holomorphic forms} 

The function $F(z,\tilde{z},\tilde{\tilde{z}})$ is called 
holomorphic (or double analytic) 
if
\beqa
\label{holo}
\frac{\partial F(z,\tilde{z},\tilde{\tilde{z}})}{\partial \tilde{z}}=\frac{%
\partial F(z,\tilde{z},\tilde{\tilde{z}})}{\partial \tilde{\tilde{z}}}=0\,. 
\eeqa
It was proved in \cite{mc2} that these conditions are equivalent to
the equations of the Cauchy-Riemann type (see also \cite{devisme1} for
analogous relations)

\begin{eqnarray}\label{holo2}
&&\frac{\partial f_{0}}{\partial x_{0}}=\frac{\partial f_{1}}{\partial x_{1}}%
=\frac{\partial f_{2}}{\partial x_{2}}\,,  \nonumber \\
&&\frac{\partial f_{0}}{\partial x_{1}}=\frac{\partial f_{1}}{\partial x_{2}}%
=\frac{\partial f_{2}}{\partial x_{0}}\,,   \\
&&\frac{\partial f_{0}}{\partial x_{2}}=\frac{\partial f_{1}}{\partial x_{0}}%
=\frac{\partial f_{2}}{\partial x_{1}}\,,  \nonumber 
\end{eqnarray}

\noindent
assuming that the function $F$ is derivable.
As a direct consequence, $F$ is holomorphic {\it if and only if}
the one-form  
$\omega ^{(1,0,0)}=Fdz$ is closed {\it i.e.}
$
d \omega^{(1,0,0)}= 0.$
Note, that in the polar form $dz=z(\frac{d\rho }{\rho }+d\varphi _{1}\
q+d\varphi _{2}\ q^{2})$ introducing the  function $zF(z)=h_{0}(\rho
,\varphi _{1},\varphi _{2})+h_{1}(\rho ,\varphi _{1},\varphi
_{2})q+h_{2}(\rho ,\varphi _{1},\varphi _{2})q^{2}$ the condition
$d (F(z) \ dz)=0$ gives

\beqa
\label{holo3}
\begin{array}{lll}
\frac{\partial h_{1}}{\partial \varphi _{1}}=\frac{\partial h_{2}}{\partial
\varphi _{2}}\,, & \rho \frac{\partial h_{2}}{\partial \rho }=\frac{\partial
h_{0}}{\partial \varphi _{1}}\,, & \frac{\partial h_{0}}{\partial \varphi
_{2}}=\rho \frac{\partial h_{1}}{\partial \rho }\,, \\ 
\frac{\partial h_{2}}{\partial \varphi _{1}}=\frac{\partial h_{0}\,}{%
\partial \varphi _{2}}, & \rho \frac{\partial h_{0}}{\partial \rho }=\frac{%
\partial h_{1}}{\partial \varphi _{1}}\,, & \frac{\partial h_{1}}{\partial
\varphi _{2}}=\rho \frac{\partial h_{2}}{\partial \rho }\,, \\ 
\frac{\partial h_{0}}{\partial \varphi _{1}}=\frac{\partial h_{1}}{\partial
\varphi _{2}}\,, & \rho \frac{\partial h_{1}}{\partial \rho }=\frac{\partial
h_{2}}{\partial \varphi _{1}}\,, & \frac{\partial h_{2}}{\partial \varphi
_{2}}=\rho \frac{\partial h_{0}}{\partial \rho }\,.
\end{array}
\eeqa
\medskip

The above equations \eqref{holo2},\eqref{holo3} define ternary harmonical functions similar to the
Cauchy-Riemann (Darbou-Euler) definition of the holomorphic functions in
the binary case.
From holomorphicity constraints one can derive that the three functions 
$f_0, f_1, f_2$ (see \eqref{F})
entering in the ternary double analytic function 
satisfy the ternary Laplace equation

\begin{eqnarray}
\nabla ^{3}f_{i} &=&\frac{\partial ^{3}f_{i}}{{\partial z}{\partial \tilde{z}%
}{\partial \tilde{\tilde{z}}}}=\frac{\partial ^{3}f_{i}}{\partial x_{0}^{3}}+%
\frac{\partial ^{3}f_{i}}{\partial x_{1}^{3}}+\frac{\partial ^{3}f_{i}}{%
\partial x_{2}^{3}}-3\frac{\partial ^{3}f_{i}}{\partial x_{0}\partial
x_{1}\partial x_{2}}=0\,,i=1,2,3\,.
\end{eqnarray}

\subsubsection{Second type of holomorphicity}
The second  type of holomorphicity, corresponds to the
single analyticity that is to a function $F(z,\tilde z, {\tilde {\tilde z}})$
such that $
\frac{\partial F(z,\tilde{z},\tilde{\tilde{z}})}{\partial z}=0 \,. 
$ A direct calculation gives the corresponding Cauchy-Riemann equations 

\begin{eqnarray}
\label{holo4}
&&\frac{\partial f_{0}}{\partial x_{0}}+\frac{\partial f_{1}}{\partial x_{1}}%
+\frac{\partial f_{2}}{\partial x_{2}}=0\,,  \nonumber \\
&&\frac{\partial f_{0}}{\partial x_{2}}+\frac{\partial f_{1}}{\partial x_{0}}%
+\frac{\partial f_{2}}{\partial x_{1}}=0\,,   \\
&&\frac{\partial f_{0}}{\partial x_{1}}+\frac{\partial f_{1}}{\partial x_{2}}%
+\frac{\partial f_{2}}{\partial x_{0}}=0\,,\nonumber 
\end{eqnarray}

\noindent
with the notations given in \eqref{F}. Manipulating  equations \eqref{holo4},
we get 

\[
\frac{\partial }{\partial x_{i}}\biggl \{\frac{\partial ^{3}}{\partial
x_{0}^{3}}+\frac{\partial ^{3}}{\partial x_{1}^{3}}+\frac{\partial ^{3}}{%
\partial x_{2}^{3}}-3\frac{\partial ^{3}}{\partial x_{0}\partial
x_{1}\partial x_{2}}\biggl \}f_{j}=0\,, i,j=0,1,2. 
\]
This means that

\[
\biggl \{\frac{\partial ^{3}}{\partial x_{0}^{3}}+\frac{\partial ^{3}}{%
\partial x_{1}^{3}}+\frac{\partial ^{3}}{\partial x_{2}^{3}}-3\frac{\partial
^{3}}{\partial x_{0}\partial x_{1}\partial x_{2}}\biggl \}f_{i}=C_{i}\,,
i=0,1,2 
\]
where $C_{i}$ are a constant. 
 The solution of the above
equations is the sum of the general solution of the homogeneous equations
and a particular solution of the inhomogeneous equations. The solution of
the inhomogeneous equations can be searched as a linear combination of the
functions $x_{0}^{3},\,x_{1}^{3},\,x_{2}^{3}$ and $x_{0}x_{1}x_{2}$, but it
is easy to verify, that this combination does not satisfy the above set of
equations for $f_{i}$ linear in partial derivatives $\partial /(\partial
x_{j})$. Therefore $C_{i}=0, i=0,1,2$.

In addition to these Cauchy-Riemann equations \eqref{holo4}
the holomorphic functions has to be real. This means that
$F(\tilde z, {\tilde {\tilde z}})= F(\tilde z {\tilde {\tilde z}})$ and
gives the relation

\[
\tilde z \frac{\partial}{\partial \tilde z} F(\tilde z, {\tilde {\tilde z}})=
{\tilde {\tilde z}} \frac{\partial}{\partial {\tilde {\tilde z}}}
 F(\tilde z, {\tilde {\tilde z}}).
\]

\noindent
In components this gives the additional relations

\beqa
\label{holo5}
\left(x_0 \frac{\partial }{\partial x_2} +
x_1\frac{\partial }{\partial x_1} +
x_2 \frac{\partial }{\partial x_2}\right) f_2&=&
\left(x_0 \frac{\partial }{\partial x_1} +
x_1\frac{\partial }{\partial x_0} +
x_2 \frac{\partial }{\partial x_0}\right) f_1\,, \nonumber \\
\left(x_0 \frac{\partial }{\partial x_2} +
x_1\frac{\partial }{\partial x_1} +
x_2 \frac{\partial }{\partial x_2}\right) f_0&=&
\left(x_0 \frac{\partial }{\partial x_1} +
x_1\frac{\partial }{\partial x_0} +
x_2 \frac{\partial }{\partial x_0}\right) f_2\,,  \\
\left(x_0 \frac{\partial }{\partial x_2} +
x_1\frac{\partial }{\partial x_1} +
x_2 \frac{\partial }{\partial x_2}\right) f_1&=&
\left(x_0 \frac{\partial }{\partial x_1} +
x_1\frac{\partial }{\partial x_0} +
x_2 \frac{\partial }{\partial x_0}\right) f_0 \nonumber \,.  
\eeqa

\noindent
As an example one sees that the function $F(\tilde z,{\tilde {\tilde z}})=
(\tilde z{\tilde {\tilde z}})^n$ satisfies the conditions
\eqref{holo4}-\eqref{holo5}, although the function
$F(\tilde z,{\tilde {\tilde z}})=
(\tilde z)^n({\tilde {\tilde z}})^m, n \ne m$ only satisfies
\eqref{holo4}.
\subsection{Ternary conformal transformations}

The transformations of a domain in the complex plane $\mathbb {C}$ 
into the other
domain in $\mathbb {C}$ can be performed with the use of two functions, 
\begin{eqnarray}
u =u(x,y)\,, \  
v =v(x,y)\,,  
&&
\end{eqnarray}
which are respectively the real and imaginary parts of the holomorphic
function $w=u+iv=f(z)=f(x+iy)$. The Jacobian of this transformation is: 
\[
\frac{\partial (u,v)}{\partial (x,y)}=|f^{\prime }(z)|^{2}. 
\]
If $f^{\prime }(z)\neq 0$, the corresponding transformation $w=f(z)$ is
conformal.
For the double holomorphic ternary function $F(z)=f_{0}+qf_{1}+f_{2}q^{2}$ ,
$\tilde{\partial }F=\tilde{\tilde{\partial }}F=0$,
$F$ defines a conformal transformation.  Using the Cauchy-Riemann equations
with $F'(z)=\frac{d F(z)}{dz}$, we get   

\[
\frac{\partial (f_{0},f_{1},f_{2})}{\partial (x_{0},x_{1},x_{2})}%
=|F'(z)|^{3}\,.
\]
\subsection{The Integration of the Differential Forms}

The ternary differential 1-forms can be integrated along some curves as it
was in the ordinary holomorphic case. If the form is closed the result
depends only on the homotopy class of the curve ${L}$. In particular
on the simply connected manifold (surface) such integral $\int_{L}\omega $
is a well defined function, the integral from the exact differential form
depends only on the initial and final points of the curve $L$. Moreover,
locally for each closed form one can build the exact form.

\subsubsection{Integral of holomorphic functions of the first type}

Now we want to consider the integral from the ternary holomorphic function
of the first type: 
\[
\int_L \omega = \int_{L}F(z)dz=\int_{L}(\omega _{0} +\omega _{1}q+\omega
_{2} q^2) \,, 
\]
where three associated 1-forms are given below (see \eqref{F}) 
\begin{eqnarray}
&&\omega _{0}=(f_{0}dx_{0}+f_{1}dx_{2}+f_{2}dx_{1})\,,  \nonumber \\
&&\omega _{1}=(f_{1}dx_{0}+f_{0}dx_{1}+f_{2}dx_{2})\,,  \nonumber \\
&&\omega _{2}=(f_{2}dx_{0}+f_{1}dx_{1}+f_{0}dx_{2})\,.  \nonumber \\
&&
\end{eqnarray}

\noi
It is easy to prove that  this 1-form $\omega =Fdz$ is exact for 
an  holomorphic function $F(z)$ (we have already proved that is closed). 
Indeed, introduce a zero-form
$V=V_{0}+V_{1}q+V_{2}q^{2}$, for which we assume $dV=\omega $.
This gives

\begin{eqnarray}
\label{V}
&&f_{0}=\frac{\partial V_{0}}{\partial x_{0}},\qquad f_{2}=\frac{\partial
V_{0}}{\partial x_{1}},\qquad f_{1}=\frac{\partial V_{0}}{\partial x_{2}}, 
\nonumber \\
&&f_{1}=\frac{\partial V_{1}}{\partial x_{0}},\qquad f_{0}=\frac{\partial
V_{1}}{\partial x_{1}},\qquad f_{2}=\frac{\partial V_{1}}{\partial x_{2}}, 
 \\
&&f_{2}=\frac{\partial V_{2}}{\partial x_{0}},\qquad f_{1}=\frac{\partial
V_{2}}{\partial x_{1}},\qquad f_{0}=\frac{\partial V_{2}}{\partial x_{2}}, 
\nonumber 
\end{eqnarray}

\noindent
these equations admit a solution  {\it if and only if} the function
$V$ is holomophic because 
\eqref{V} are just the Cauchy-Riemann  equations \eqref{holo2}
for $V$. 
Therefore, if $V$ is holomorphic, we have

\begin{eqnarray}
\int_{L}F(z)dz = V(z_1) -V(z_2)
\end{eqnarray}

\noi
with $z_1$ and $z_2$ being the end points of the curve $L$.
\medskip

Now we consider an important example of  ternary holomorphic functions

\[
\ln \,z=\int^{z}\frac{dz^{\prime }}{z^{\prime }}=(\ln z)_{0}+(\ln
z)_{1}q+(\ln z)_{2}q^{2}\,,
\]
Of course the curve along which we perform the integration is such that
$1/z$ is always defined, {\it i.e.} there is no singular ternary
complex numbers.
In the polar form this integral is easily calculated

\beqa
\label{polar2}
\ln z=\int^{z}\frac{dz^{\prime }}{z^{\prime }}=\int^{z}\left( \frac{d\rho
^{\prime }}{\rho ^{\prime }}+d\varphi _{1}^{\prime }q+d\varphi _{2}^{\prime
}q^{2}\right) =\ln \rho +\varphi _{1}q+\varphi _{2}q^{2}\,.
\eeqa
This result corresponds to the obvious relation 
\[
e^{\ln z}=z\,. 
\]

It is useful to perform the calculation of the integral directly in
the ``cartesian coordinates''.
One can take easily the  integral for $(\ln z)_0$ 
\begin{eqnarray}
\label{rho}
\ln \rho=(\ln z)_{0} &=&\int \frac{1}{|z|^{3}}[%
(x_{0}^{2}-x_{1}x_{2})dx_{0}+(x_{1}^{2}-x_{0}x_{2})dx_{1}+
(x_{2}^2-x_{0}x_{1})dx_{2})%
]  \nonumber \\
&=&\frac{1}{3}\ln (x_{0}^{3}+x_{1}^{3}+x_{2}^{3}-3x_{0}x_{1}x_{2}).
\end{eqnarray}

\noi
However, the  second and third integrals  are more involved
\[
(\ln z)_{1}=\int \frac{1}{|z|^{3}}[%
(x_{2}^{2}-x_{0}x_{1})dx_{0}+(x_{0}^{2}-x_{1}x_{2})dx_{1}+(x_{1}^{2}-x_{0}x_{2})dx_{2}%
]\,, 
\]
\[
(\ln z)_{2}=\int \frac{1}{|z|^{3}}[%
(x_{1}^{2}-x_{0}x_{2})dx_{0}+(x_{2}^{2}-x_{0}x_{1})dx_{1}+(x_{0}^{2}-x_{1}x_{2})dx_{2}%
]\,. 
\]

\noi
If we denote $\varphi_1=(\ln z)_{1}=\int \frac{1}{|z|^{3}}
(a dx_0 + b dx_1 + c dx_2)$ a direct verification gives
$\frac{\partial b}{\partial x_0}= \frac{\partial a}{\partial x_1}$
plus similar relations for the other terms (this was expected see \eqref{polar2}). 
Thus, we have 

\[
\frac{\partial \varphi _{1}}{\partial x_{0}}=
\frac{x_{2}^{2}-x_{0}x_{1}}{|z|^{3}%
},\frac{\partial \varphi _{1}}{\partial x_{1}}=\frac{x_{0}^{2}-x_{1}x_{2}}{%
|z|^{3}},\frac{\partial \varphi _{1}}{\partial x_{2}}=\frac{x_{1}^{2}-x_{0}x_{2}%
}{|z|^{3}},
\]

\noi
and

\begin{eqnarray}
\label{phi1}
\varphi _{1}=(\ln z)_{1}&=&\int \frac{dx_{0}(x_{2}^{2}-x_{0}x_{1})}{%
(x_{0}+x_{1}+x_{2})(x_{0}+jx_{1}+j^{2}x_{2})(x_{0}+j^{2}x_{1}+jx_{2})} 
 \\
&=&\frac{1}{3}[\ln (x_{0}+x_{1}+x_{2})+j^{2}\ln
(x_{0}+jx_{1}+j^{2}x_{2})+j\ln (x_{0}+j^{2}x_{1}+jx_{2})] \,.  \nonumber
\end{eqnarray}

\noi
Similarly, 
\begin{eqnarray}
\label{phi2}
\varphi _{2} &=&(\ln z)_{2}=\frac{1}{3}[\ln (x_{0}+x_{1}+x_{2})+j\ln
(x_{0}+jx_{1}+j^{2}x_{2})+j^{2}\ln (x_{0}+j^{2}x_{1}+jx_{2})]\,.  \nonumber \\
\end{eqnarray}

\noi
{\it A priori} these functions seem to be ill-defined
(multi-valued)  and cuts in the
complex plane should be taken, however it is not the case if $\rho >0$
(if $\rho \le 0$ the formul\ae \ above are not defined). Indeed, one can write

\[
\varphi_1= -\frac16 \ln [(x_0 +jx_1 +j^2 x_2)(x_0 + j^2 x_1 + j x_2)] +
\frac{2}{\sqrt 3} \psi
\]
with 

\[
\psi= 
\mathrm{arctg}[\frac{\sqrt{3}(x_{1}-x_{2})}{2x_{0}-x_{1}-x_{2})}]\,, 
\]
and similar relations for $\varphi_2$.
Using above relations we obtain finally

\begin{eqnarray}
(\ln z)&=&  
\frac{1}{3}\ln |z|^{3}   \\
&+&q[\frac{1}{3}[\ln (x_{0}+x_{1}+x_{2})+j^{2}\ln
(x_{0}+jx_{1}+j^{2}x_{2})+j\ln (x_{0}+j^{2}x_{1}+jx_{2})]  \nonumber \\
&+&q^{2}\frac{1}{3}[\ln (x_{0}+x_{1}+x_{2})+j\ln
(x_{0}+jx_{1}+j^{2}x_{2})+j^{2}\ln (x_{0}+j^{2}x_{1}+jx_{2})]\,.  \nonumber
\end{eqnarray}

\noi
It is interesting to notice that adding $\ln z$ with its ternary conjugations 
one gets

\[
\ln z+\ln \tilde{z}+\ln \tilde{\tilde{z}}=\ln |z|^{3}. 
\]

Let us now  consider $z$ in the ``polar'' coordinates (see \eqref{EKI})

\[
z=\exp \{\ln z\}=\rho \exp \{\varphi _{1}q+\varphi _{2}q^{2}\}=\rho \exp
\{(2K_{0}-E_{0})\varphi +I\sqrt{3}\theta \}, 
\]
where

\[
\ln \rho =\frac13(\ln[%
(x_{0}+x_{1}q+x_{2}q^{2})(x_{0}+jx_{1}+j^{2}x_{2})(x_{0}+j^{2}x_{1}+jx_{2})%
]), 
\]

\beqa  \label{non-compact}
\varphi&=&\frac{\varphi_1+\varphi_2}{2}=\frac{(\ln z)_1+(\ln z)_2}{2} =\frac{1}{2}%
\ln [\frac{x_0+x_1+x_2}{\rho}], \\
\theta& =&\frac{\varphi _{1}-\varphi _{2}}{2}=\frac{(\ln z)_{1}-(\ln z)_{2}}{2}=%
\frac{i}{2\sqrt{3}}\ln[\frac{x_{0}+jx_{1}+j^{2}x_{2}}{x_{0}+j^{2}x_1+jx_{2}}%
]=\frac{1}{\sqrt{3}}\psi.  \nonumber
\eeqa

The relations, \eqref{rho}, \eqref{phi1} and \eqref{phi2} which
give the polar coordinates as functions of the cartesian ones, can be
considered as inverse formul\ae\ for the multisine-functions \eqref{mus2},
which present the cartesian coordinates as functions of polar ones
(these are well defined since by definition the angle in the compact
direction is restricted to $[0, 2 \pi /\sqrt{3}[$ see \eqref{polar}). As a
matter of a formal exercise, one can check that the expressions %
\eqref{non-compact} inserted in the multisine-functions %
\eqref{mus2} give $x_{k}=\rho m_{k}(\varphi _{1},\varphi _{2})$.

One can also verify the relation 
\[
\int^{z}\frac{dz}{z}=\int \frac{d\rho }{\rho }+(2K_{0}-E_{0})\int
d\varphi +\sqrt{3}I\int d\theta \,. 
\]
In particular on the cubic surfaces, where $\rho =\text{const},$
 $\varphi =\text{const}$,
{\it i.e.} if we integrate  over the closed contour around the
``trisectrice'',
the integral  equals $2\pi I$. This result can also be 
obtained using the parametrisation
given in  \eqref{trumpet}.

\subsubsection{Integral of holomorphic functions of the second type}

Let us consider now the integral of holomorpic functions of the
second  type, which is related to the
single ternary holomorphicity. We perform the integration of the two-form
over a surface $S$ with  some boundary $\partial S$ 
\begin{eqnarray}
\int_S \Omega = \int_{S}\Phi (\tilde z {\tilde {\tilde z}})\frac{d\tilde{z}\wedge d\tilde{\tilde{z}}}{j^2-j}
&=&\int_{S}(\Omega _{0}+\Omega _{1}q+
\Omega_{2}q^2)\,.  
\end{eqnarray}
Here from $\Phi(\tilde z {\tilde {\tilde z}})=\Phi_0 + \Phi_1 q + \Phi_2 q^2$, 
we defined three associated 2-forms 
\begin{eqnarray}
&&\Omega _{0}=\Phi _{0}dx_{1}\wedge dx_{2}+\Phi _{1}dx_{2}\wedge
dx_{0}+\Phi _{2}dx_{0}\wedge dx_{1}\,,  \nonumber \\
&&\Omega _{1}=\Phi _{1}dx_{1}\wedge dx_{2}+\Phi _{2}dx_{2}\wedge
dx_{0}+\Phi _{0}dx_{0}\wedge dx_{1} \,,  \\
&&\Omega _{2}=\Phi _{2}dx_{1}\wedge dx_{2}+\Phi _{0}dx_{2}\wedge
dx_{0}+\Phi _{1}dx_{0}\wedge dx_{1} \,.  \nonumber 
\end{eqnarray}

This integral vanishes on a closed surface if $\Omega = d \omega$. If
$\omega = \omega_z dz + \omega_{\tilde z} d \tilde z + \omega_{{\tilde {\tilde z}}} d {\tilde {\tilde z}}$,
with (i) $\omega_{z} $ being a
holomorphic function of the first first type and 
(ii) $\omega_{\tilde z}$ and $ \omega_{{\tilde {\tilde z}}}$ 
being holomorphic functions of the second  type,
satisfying  the following constraints
$\partial_{\tilde z} \omega_{{\tilde {\tilde z}}} -
\partial_{{\tilde {\tilde z}}} \omega_{\tilde z} = \frac{1}{j^2 -j} \Phi(\tilde z {\tilde {\tilde z}})$,
we have $\Omega = d \omega$.

Let us consider the following example, where the function $\Phi $ has
singularities

\beqa
\label{mono}
 \int \Omega =\frac{1}{j^2 -j} \int \frac{d\tilde{z}\wedge d\tilde{\tilde{z}}}{%
(\tilde{z}\tilde{\tilde{z}})}=\frac{1}{j^2 -j}\int \frac{zd\tilde{z}\wedge d{\tilde
{\tilde{z}}}}{\rho^3}\,,
\eeqa

\begin{eqnarray}
\Omega =\frac{1}{j^{2}-j}\frac{z}{\rho^3}d\tilde{z}\wedge d\tilde{\tilde{z}}
&=&\frac{1}{\rho^3}(x_{0}dx_{1}\wedge dx_{2}+x_{1}dx_{2}\wedge dx_{0}+x_{2}dx_{0}\wedge
dx_{1})  \nonumber \\
&+&\frac{1}{\rho^3}(x_{0}dx_{0}\wedge dx_{1}+x_{1}dx_{1}\wedge dx_{2}+x_{2}dx_{2}\wedge
dx_{0})q  \\
&+&\frac{1}{\rho^3}(x_{0}dx_{2}\wedge dx_{0}+x_{1}dx_{0}\wedge dx_{1}+x_{2}dx_{1}\wedge
dx_{2})q^2\,.  \nonumber 
\end{eqnarray}

To integrate $\Omega$ over a piece $\Sigma $ of the three-dimensional
surface, $z\tilde{z}\tilde{\tilde{z}}=\rho ^{3}$, one should choose a
parametrization, $g:\mathcal{R}\rightarrow \Sigma $ , where $\mathcal{R}$ is
a subset of ${\mathbb R}^{2}$: 
\[
g(u,v)=(x_{0}(u,v),x_{1}(u,v),x_{2}(u,v))\,, 
\]
which gives $dx_i \wedge d x_j= J_{ij} du dv$, where the Jacobians are 

\[
J_{ij}= 
\begin{array}{|cc|}
\frac{\partial x_{i}}{\partial u} & \frac{\partial x_{j}}{\partial u} \\ 
\frac{\partial x_{i}}{\partial v} & \frac{\partial x_{j}}{\partial v}
\end{array}
\,, \, i,j=0,1,2\,. 
\]
If we choose the parametrisation \eqref{trumpet}, with $a =x_0+x_1+x_2$ and $a r^2 = \rho^3$ we get
(considering here the case where $a>0$ corresponding to $\rho>0$).

\begin{eqnarray}
x_{0} &=&\frac{a}{3}-\frac{2}{3}\frac{\rho ^{3/2}}{\sqrt{a}}\cos \theta \,, 
\nonumber \\
x_{1} &=&\frac{a}{3}+\frac{1}{3}\frac{\rho ^{3/2}}{\sqrt{a}}(\cos \theta +%
\sqrt{3}\sin \theta ) \,,  \nonumber \\
x_{2} &=&\frac{a}{3}+\frac{1}{3}\frac{\rho ^{3/2}}{\sqrt{a}}(\cos \theta -%
\sqrt{3}\sin \theta )\,.  \nonumber \\
&&
\end{eqnarray}
The region of integration is given below 
\[
0<a_{1}\leq a\leq a_{2},\qquad 0\leq \theta \leq 2\pi . 
\]
We obtain also the relations 
\begin{eqnarray}
x_{0}^{2}+x_{1}^{2}+x_{2}^{2} &=&\frac{a^{2}}{3}\biggl (1+2\frac{\rho ^{3}}{%
a^{3}}\biggl )\,,  \nonumber \\
x_{0}x_{1}+x_{1}x_{2}+x_{2}x_{0} &=&\frac{a^{2}}{3}\biggl (1-\frac{\rho ^{3}%
}{a^{3}}\biggl )\,.  \nonumber \\
&&
\end{eqnarray}

\noi
Based on the above parametrization one can calculate the Jacobians:

\begin{eqnarray}
J_{12} &=&\frac{\sqrt{3}}{9}\frac{\rho ^{3}}{a^{2}}-\frac{2\sqrt{3}}{9}\frac{%
\rho ^{3/2}}{\sqrt{a}}\biggl(\cos \theta \biggl )\,,  \nonumber \\
J_{20} &=&\frac{\sqrt{3}}{9}\frac{\rho ^{3}}{a^{2}}+\frac{\sqrt{3}}{9}\frac{%
\rho ^{3/2}}{\sqrt{a}}\biggl(\cos \theta +\sqrt{3}\sin \theta \biggl )\,, 
 \\
J_{01} &=&\frac{\sqrt{3}}{9}\frac{\rho ^{3}}{a^{2}}+\frac{\sqrt{3}}{9}\frac{%
\rho ^{3/2}}{\sqrt{a}}\biggl(\cos \theta -\sqrt{3}\sin \theta \biggl )\,, 
\nonumber 
\end{eqnarray}

Note that the geometrical meaning of $J_{01},J_{12},J_{20}$ is the surface
on the planes $\{01\}$, $\{12\}$, $\{20\}$, respectively. Therefore one  can
consider the following cubic relation for $J_{jk}$ as a ternary analog of the binary
Pythagor theorem
\[
J_{01}^{3}+J_{12}^{3}+J_{20}^{3}-3J_{01}J_{12}J_{20}=\frac{1}{3\sqrt{3}}%
\frac{\rho ^{6}}{a^{3}}\,. 
\]
Whereas the binary Pythagor theorem relates the lengths of the sides in a
triangle, the ternary Pythagor theorem gives a relation among the squares of
the faces, $A,B,C,D$, in the tetrahedron:

\[
S_{A}^{3}+S_{B}^{3}+S_{C}^{3}-3S_{A}S_{B}S_{C}=S_{D}^{3}\,, 
\]
where $S_i$ is the square of the corresponding face.

To obtain the final result of integration one can take into account the
relations

\begin{eqnarray}
&&x_{0}J_{12}+x_{1}J_{20}+x_{2}J_{12}=\frac{1}{\sqrt{3}}\frac{\rho ^{3}}{a}%
\,,  \nonumber \\
&&x_{0}J_{01}+x_{1}J_{12}+x_{2}J_{20}=0\,,  \nonumber \\
&&x_{0}J_{20}+x_{1}J_{01}+x_{2}J_{12}=0\,,  \nonumber \\
&&
\end{eqnarray}

\begin{eqnarray}
\int \Omega&=&
=\int  (x_0J_{12}+x_1J_{20}+x_2J_{01})\frac{a}{\rho^3} d \theta da = 
 \frac{2 \pi}{\sqrt{3}} \ln \frac{a_2}{a_1}\,.  
\end{eqnarray}

We can perform the calculation in the polar coordinates. In this case one
obtains

\[
\Omega =\frac{1}{j^2-j}\frac{d\tilde{z}\wedge d{\tilde{\tilde{z}}}}{\tilde{%
z}{\tilde{\tilde{z}}}}=d\varphi _{1}\wedge d\varphi _{2}+\frac{d\rho}{\rho}\wedge
d\varphi _{1}\ q+d\varphi _{2}\wedge \frac{d\rho}{\rho}\ q^{2}\, 
\]

\noindent for the cubic surface $\Sigma $ parametrized by $\varphi ^{\prime
}\leq \varphi \leq \varphi ^{\prime \prime },0\leq \theta \leq 2\pi/\sqrt{3} $. Using 
$\varphi _{1}=\varphi +\theta ,\varphi _{2}=\varphi -\theta $, we get 
\[
\int_{\Sigma }\Omega =\int_{\Sigma }d\varphi _{1}\wedge d\varphi _{2}=
2 \int\limits_{0}^{2\pi/\sqrt{3} }
d\theta \int\limits_{\varphi ^{\prime
}}^{\varphi ^{\prime \prime }}d\varphi =\frac{4\pi }{\sqrt{3}}(\varphi
^{\prime \prime }-\varphi ^{\prime })\,. 
\]
This relation coincides with the above result since $a=\rho e^{2 \varphi}$ 
(see \eqref{mus2}).

Let us consider now a generic two-form $\Omega $. It will be exact if there
exists a 1-form $\omega $ (not necessarily holomorphic having $d\omega = \Omega$%
). In particular for the 1-form given below

\[
\omega =\varphi _{1}d\varphi _{2}+\ln \rho \,d\varphi _{1}q+\varphi _{2}%
\frac{d\rho }{\rho }\ q^{2} 
\]
we obtain

\[
d \omega = \left(d \rho \frac{\partial}{\partial \rho} + d \varphi_1 \frac{%
\partial }{\partial \varphi_1}+ d \varphi_2 \frac{\partial }{\partial
\varphi_2}\right) \wedge \omega= \frac{1}{j^2-j}\frac{d \tilde z \wedge d {%
\tilde {\tilde z}}}{\tilde z {\tilde {\tilde z}}}\,. 
\]

\noindent Thus, $\oint_{\Sigma }\frac{1}{j-j^{2}}\frac{d\tilde{z}\wedge d{%
\tilde{\tilde{z}}}}{\tilde{z}{\tilde{\tilde{z}}}}=0$ for any closed surface
without inner singularity points.

\subsection{Physical interpretation of the singular ternary potential}

As we have seen, the integral \eqref{mono}
over the closed surface is zero if the
singularity at the trisectrice $x_{1}=x_{2}=x_{0}$ is not inside this
surface. We interprete the integrand as a magnetic field
\begin{equation}
\vec{H}=\frac{\vec{x}}{|z|^{3}}\,,
\end{equation}
because due to the ternary holomorphicity of the second type (the single
analyticity) the relation
\begin{equation}
\vec{\nabla}\vec{H}=0
\end{equation}
is valid everywhere apart from the singularity. Note, that the transmutation
of field components $\vec{H}=(H_{1,}H_{2},H_{0})\rightarrow \vec{H}^{\prime
}=(H_{2,}H_{0},H_{1})$ is equivalent to the rotation of the vector $\vec{H}$
around the trisectrice on the angle $2\pi /3$. The new vector $\vec{%
H}^{\prime }$ satisfies also the condition $\vec{\nabla}\vec{H}^{\prime }=0$
in the region outside the trisectrice. It is convenient to rotate the
coordinate system
\begin{equation}\label{rot1}
\vec{n}_{0}=\frac{1}{\sqrt{3}}\,(1,1,1)\,,\,\,\vec{n}_{1}=\frac{1}{\sqrt{2}}%
\,(1,-1,0)\,,\,\,\vec{n}_{2}=\frac{1}{\sqrt{6}}\,(1,1,-2)\,.
\end{equation}  
in such way, that the new coordinates are
\begin{eqnarray}\label{rot2}
r_{0} &=&\frac{x_{1}+x_{2}+x_{0}}{\sqrt{3}}\,,\,\,r_{1}=\frac{x_{1}-x_{2}}{%
\sqrt{2}}\,,\,\,r_{2}=\frac{x_{1}+x_{2}-2x_{0}}{\sqrt{6}}  \nonumber \\
&&
\end{eqnarray}
and the new components of the magnetic field are
\begin{equation}
\vec{h}=\frac{3\sqrt{3}}{2}\vec{H}=\frac{\overrightarrow{R}}{l|\vec{r}|^{2}}%
\,,\,\,R_{0}=l\,,\,R_{1,2}=r_{1,2}\,,\,\,\,|\vec{r}|^{2}=r_{1}^{2}+r_{2}^{2}%
\,.
\end{equation}

Let us integrate $\vec{\nabla}\overrightarrow{h}$ over the inner part of the
cylinder of the radius $a$ with the coordinate axes $l$ situated in its
center. We use the Stokes theorem to express this volume integral through
the integral over the cylinder surface surrounding the trisectrice
\begin{equation}
\int d^{2}r\,dl\,\,\vec{\nabla}\overrightarrow{h}=\int_{-\infty }^{\infty }%
\frac{dl}{l}\,2\pi \,
\end{equation}
independently of its radius $a$. It means, that the correct equation for $%
\overrightarrow{h}$ is
\begin{equation}
\,\vec{\nabla}\overrightarrow{h}=4\pi \rho \,,\,\,\rho =\frac{1}{2}\,\frac{1%
}{l}\,\delta ^{2}(r)\,,
\end{equation}
where $\rho $ is the magnetic monopole density and $\overrightarrow{r}$ is   
the coordinate transversal to the vector $\vec{n}_{0}$. The monopole string
with an infinitesimal small radius is stretched along the vector $\vec{n}%
_{0} $. We can calculate the density $\rho $ directly from the expression
for $\overrightarrow{h}$ using the relations

\begin{equation}
\vec{\nabla}_2^{2}\ln \,|r|^{2}=\overrightarrow{\partial }\frac{2%
\overrightarrow{r}}{|r|^{2}}=4\pi \delta ^{2}(r)\,.
\end{equation}

Let us find the scalar potential corresponding to this magnetic density. It
satisfies the equation
\begin{equation}
\nabla_3^{2}\varphi =\rho \,.
\end{equation}
Using the well known relation
\begin{equation}
\nabla_3^{2}\frac{1}{\left| \overrightarrow{R}\right| }=-4\pi
\,\delta ^{3}(R)\,,\,\,\overrightarrow{R}=l\overrightarrow{n}_{0}+%
\overrightarrow{r}\,,\,\,\left| \overrightarrow{R}\right| ^{2}=l^{2}+|r|^{2}
\end{equation}
we obtain for the scalar potential
\begin{equation}
\varphi (l,\overrightarrow{r})=-\frac{1}{2}\int_{-\infty }^{\infty }\frac{%
dl^{\prime }}{l^{\prime }}\frac{1}{\sqrt{(l-l^{\prime })^{2}+|r|^{2}}}=\frac{%
1}{2\left| \overrightarrow{R}\right| }\,\ln \frac{\left| \overrightarrow{R}%
\right| -l}{\left| \overrightarrow{R}\right| +l}\,.
\end{equation}

One can easily calculate the magnetic field which corresponds to this
density
\[
h_{0}^{pot}=\partial _{0}\varphi =-\frac{l}{2\left| \overrightarrow{R}%
\right| ^{3}}\,\ln \frac{\left| \overrightarrow{R}\right| -l}{\left|
\overrightarrow{R}\right| +l}-\frac{1}{\left| \overrightarrow{R}\right| ^{2}}%
\,,
\]
\begin{equation}
\,\overrightarrow{h}^{pot}=\overrightarrow{\partial }\varphi =-\frac{%
\overrightarrow{r}}{2\left| \overrightarrow{R}\right| ^{3}}\,\ln \frac{%
\left| \overrightarrow{R}\right| -l}{\left| \overrightarrow{R}\right| +l}+%
\frac{l\,\overrightarrow{r}}{|r|^{2}\left| \overrightarrow{R}\right| ^{2}}\,.
\end{equation}
This is only a potential part of the field $h_{\mu }$, which can be written 
as follows
\begin{equation}
h_{\mu }=h_{\mu }^{pot}+h_{\mu }^{rot}\,,\,\,h_{\mu }^{pot}=\frac{\partial
_{\mu }\partial _{\sigma }}{\partial ^{2}}\,h_{\sigma }\,,\,\,%
\overrightarrow{h}^{rot}=-\frac{1}{\partial ^{2}}\left[ \vec{\nabla},\left[
\vec{\nabla},\overrightarrow{h}\right] \right]
\end{equation}
and $h_{\mu }^{rot}$ is related to the existence of closed electric 
currents. By subtracting $h_{\mu }^{pot}$ from $h_{\mu }$ we obtain
\begin{equation}
h_{0}^{rot}=\frac{l}{2\left| \overrightarrow{R}\right| ^{3}}\,\ln \frac{%
\left| \overrightarrow{R}\right| -l}{\left| \overrightarrow{R}\right| +l}+%
\frac{1}{\left| \overrightarrow{R}\right| ^{2}}+\frac{1}{|r|^{2}}\,,\,\,%
\overrightarrow{h}^{rot}=\frac{\overrightarrow{r}}{2\left| \overrightarrow{R}%
\right| ^{3}}\,\ln \frac{\left| \overrightarrow{R}\right| -l}{\left|
\overrightarrow{R}\right| +l}+\frac{\overrightarrow{r}}{\left|
\overrightarrow{R}\right| ^{2}l}\,.
\end{equation}
The divergency of this vector does not contain the singularity at $\overrightarrow{r}=0$ 
\begin{equation}
\partial _{i}h_{i}^{rot}=\left( \frac{|r|^{2}-2l^{2}}{2\left|  
\overrightarrow{R}\right| ^{5}}+\frac{2l^{2}-|r|^{2}}{2\left|
\overrightarrow{R}\right| ^{5}}\right) \,\ln \frac{\left| \overrightarrow{R}%
\right| -l}{\left| \overrightarrow{R}\right| +l}-\frac{l}{\left|
\overrightarrow{R}\right| ^{4}}+\frac{l}{\left| \overrightarrow{R}\right|
^{4}}-2\frac{l}{\left| \overrightarrow{R}\right| ^{4}}+2\frac{l}{\left|
\overrightarrow{R}\right| ^{4}}=0\,.
\end{equation}
Note, however, that the vector $\vec{h}^{rot}$ does not have this property  
after the rotation around the axes $l$ on the angle $2\pi /3$ and therefore
it can not be
expressed in terms of the ternary number with the single analyticity.

\noindent
The electromagnetic current according to the Maxwell equations can be found
in terms of $\overrightarrow{h}^{rot}$ as follows
\begin{equation}
h_{0}=\frac{1}{\,|r|^{2}}\,,\,\,h_{i}=\,\frac{r_{i}}{l\,|r|^{2}}%
\,\,(i=1,2)\,.
\end{equation}
\begin{equation}
\vec{j}=[\vec{\nabla},\overrightarrow{h}]\,=[\vec{\nabla},\overrightarrow{h}%
^{rot}]\,,
\end{equation}
because the potential part does not give to it any contribution.

In an explicit form we have for our case
\begin{equation}
j_{0}=0\,,\,\,j_{k}=\epsilon _{kl}\,r_{l}\,\left( -\frac{2}{\left| r\right|
^{4}}+\frac{1}{l^{2}|r|^{2}}\,\right) \,,\,
\end{equation}
where $\epsilon _{kl}$ is the two-dimensional anti-symmetric tensor with $%
\epsilon _{12}=1$. Because $\overrightarrow{j}.\overrightarrow{r}=0$, the   
current $\overrightarrow{j}$ is tangential to the circles enclosing the
trisectrice $\vec{n}_{0}$ and have the same value on them
\begin{equation}
|j|=|r|\,\left| -\frac{2}{\left| r\right| ^{4}}+\frac{1}{l^{2}|r|^{2}}%
\right| \,.   
\end{equation}  
Their direction is changed to opposite one at the distance

\begin{equation}
|r|=\sqrt{2}\,l\,.
\end{equation}

We have the following representation for $h_{\mu }^{rot}$ in terms of the
vector-potential $\vec{A}$
\begin{equation}
\overrightarrow{h}^{rot}=[\vec{\nabla},\vec{A}]\,.
\end{equation}
It is valid everywhere including the trisectrice. Instead of this equation 
for $\vec{A}$ we consider a simpler equation valid in the region of
analyticity of $\vec{h}$
\[
\vec{h}=\frac{\vec{r}}{l|\vec{r}|^{2}}=\overrightarrow{\text{rot}}\vec{A},
\]

Choosing the gauge
\begin{eqnarray}
&&A_{0}=0\,,  \nonumber \\
&&
\end{eqnarray}
we find its solution in the form
\begin{eqnarray}
&&A_{0}=0\,,  \nonumber \\
&&A_{1}=\frac{r_{2}}{r_{1}^{2}+r_{2}^{2}}\,\ln \frac{l}{\sqrt{%
r_{1}^{2}+r_{2}^{2}}}\,,  \nonumber \\
&&\,A_{2}=-\frac{r_{1}}{r_{1}^{2}+r_{2}^{2}}\,\ln \frac{l}{\sqrt{%
r_{1}^{2}+r_{2}^{2}}}\,.  \nonumber \\
&&
\end{eqnarray}
  
For $W=\varphi_0 + \varphi_1 q + \varphi_2 q^2$, 
note, that one can consider the integration of the three form
\begin{eqnarray}
\int_{M}W\frac{d\tilde{z}\wedge d\tilde{\tilde{z}}\wedge 
d\tilde{\tilde{\tilde{z}}}}{j^2-j}
&=&\int_{M}(\omega _{0}^{(3)}+\omega _{2}^{(3)}q^{2}+\omega
_{1}^{(3)}q)\,,  \nonumber \\
&&
\end{eqnarray}

\begin{eqnarray}
\omega^{(3)}_i=\varphi_i dx_0 \wedge dx_1 \wedge dx_2.
\end{eqnarray}

In this case there is also a correspondence between the integrals written in
the vector calculus and in terms of differential forms:

\begin{eqnarray}
&&\int_M W d V \leftrightarrow \int_M \Omega_W^{(3)}  \nonumber \\
&&\int_M ( \vec \nabla \times \vec H) d V \leftrightarrow \int_M d
\Omega_W^{(3)}  \nonumber \\
\end{eqnarray}  

If $\mathit{V} $ is a volume in $\mathbb R^3$, then we have:
\begin{eqnarray}
\int_{\partial \mathit{V}} (\vec H d \vec S)= \int_{\partial \mathit{V}}
\Omega_W^{(2)}= \int_{\mathit{V}} \Omega_W^{(3)} =\int_{\mathit{V}} (\nabla
\cdot \vec H) d \mathit{V}
\end{eqnarray}  


\section{Ternary mechanics and monopole dynamics}

\subsection{Newton equation for a particle in a ternary field}

\bigskip Let us consider the ternary generalization of the classical \
Newton equation
\begin{equation}
\frac{d^{2}z}{(dt)^{2}}=-G\,\frac{1}{\widetilde{z}\,\widetilde{\widetilde{z}}%
}\,.
\end{equation}
where $G$ is a coupling constant. 
We can write the Newton equation in the vector form
\begin{equation}
\frac{d^{2}\overrightarrow{x}}{(dt)^{2}}=-G\,\frac{\overrightarrow{x}}{Z^{3}}%
\,,
\end{equation}
where $\overrightarrow{x}=(x_{1},x_{2},x_{0})$ and
\begin{equation}
Z^{3}\equiv z\,\widetilde{z}\,\widetilde{\widetilde{z}}%
=x_{1}^{3}+x_{2}^{3}+x_{0}^{3}-3\,x_{1}\,x_{2}\,x_{0}\,.
\end{equation}

We interprete the ternary Newton equation as the equation for a monopole
moving in the magnetic field $H$ of the type
\begin{equation}
\overrightarrow{H}=\frac{\overrightarrow{x}}{Z^{3}}\,,
\end{equation}
because due to the analyticity condition for the ternary function $%
H=H_{0}+qH_{1}+q^{2}H_{2}$ in an accordance with the Maxwell equations one
obtains $\overrightarrow{\nabla }\overrightarrow{H}=0$ everywhere apart the
singularities on the trisectrice $x_{1}=x_{2}=x_{0}$.
After the change of coordinates \eqref{rot1}-\eqref{rot2} the Newton equation becomes

\begin{equation}
\frac{d^{2}\overrightarrow{r}}{(dt)^{2}}=-g\,\overrightarrow{h}\,,\,\,%
\overrightarrow{h}=\frac{\overrightarrow{r}}{|r|^{2}l}\,,\,\,g=G\frac{2}{3%
\sqrt{3}}\,.
\end{equation}

Thus, we consider in fact the movement of a monopole-type object in the   
magnetic field of a simple form, created by the monopoles situated on the
line $r_{1}=r_{2}=0$

\begin{equation}
4\pi \rho =\partial _{0}h_{0}+\partial _{1}h_{1}+\partial _{2}h_{2}=\frac{%
2\pi }{|l|}\,\delta ^{2}(r)
\end{equation}
and by electromagnetic currents

\begin{equation}
j_{0}=0\,,\,\,j_{k}=\epsilon _{kl}\,r_{l}\,\left( -\frac{2}{\left| r\right|
^{4}}+\frac{1}{l^{2}|r|^{2}}\right) \,,
\end{equation}
circulated around the trisectrice. We show below, that the above ternary   
equation is integrable similar to the case of the Newton equation in the
cental gravitational field.

It is obvious, that there is an integral of motion - the angular moment   
\begin{equation}
\overrightarrow{M}=\left[ \overrightarrow{r},\partial _{t}\overrightarrow{r}%
\right] \,\,,\,\,\frac{d}{dt}\,\overrightarrow{M}=0
\end{equation}  
or in the components

\begin{equation}
M_{0}=r_{1}\,\partial _{t}r_{2}-r_{2}\,\partial
_{t}r_{1}\,\,,\,\,M_{1}=r_{2}\,\partial _{t}l-l\,\partial
_{t}r_{2}\,,\,\,M_{2}=l\,\partial _{t}r_{1}-r_{1}\,\partial _{t}l\,\,.
\end{equation}

To begin with, let us consider the particular case, in which the particle
moves in the plane to which the trisectrice belongs. Without the loss of   
generality we can put

\begin{equation}
r_{2}=M_{0}=M_{1}=0\,.
\end{equation}  
In this case the only non-trivial integral of motion is
\begin{equation}
M_{2}=l\,\partial _{t}r_{1}-r_{1}\,\partial _{t}l\,\,.
\end{equation}
This constraint should be considered together with the Newton equation for $%
r_{1}$

\begin{equation}
\partial _{t}^{2}\,r_{1}=-\frac{g}{r_{1}}\,\frac{1}{l}\,.
\end{equation}

It is convenient to introduce the new variable
\begin{equation}
z=\frac{l}{r_{1}}\,. 
\end{equation}
In the variables ($r_{1},z$) we can rewrite as follows the integral of motion
\begin{equation}
M_{2}=-r_{1}^{2}\,\partial _{t}z\,.
\end{equation}
and the Newton equation
\begin{equation}
\partial _{t}^{2}\,r_{1}=-\frac{g}{r_{1}^{2}}\,\frac{1}{z}=\frac{g}{M_{2}}\,%
\frac{\partial _{t}z}{z}
\end{equation}
One can integrate the last equation
\begin{equation}
\partial _{t}\,r_{1}=\frac{g}{M_{2}}\,\,\ln \left| \frac{z}{z_{0}}\right| ,
\end{equation}
where $z_{0}$ is the value of the ratio $l/r_{1}$ on the trajectory for
which $\partial _{t}\,r_{1}=0$.

With the use of the conserved momentum $M_{2}$ we rewrite the above integral
of motion as follows
\begin{equation}
\partial _{t}\,\sqrt{-\frac{M_{2}}{\partial _{t}z}}=-\frac{1}{2\partial _{t}z%
}\,\sqrt{-\frac{M_{2}}{\partial _{t}z}}\,\partial _{t}^{2}z=\frac{g}{M_{2}}%
\,\,\ln \left| \frac{z}{z_{0}}\right| \,
\end{equation}
and integrate it again finding the monopole trajectory
\begin{equation}
-\sqrt{-M_{2}\partial _{t}z}=\frac{M_{2}}{r_{1}}=\frac{g}{M_{2}}\,\,\left(
z\ln \left| \frac{z}{ez_{0}}\right| -z_{1}\ln \left| \frac{z_{1}}{ez_{0}}%
\right| \right) \,,
\end{equation}  
where $z_{1}$ is the value of the ratio $l/r_{1}$, for which $\partial
_{t}z=0$.

Integrating the last relation, we obtain the relation
\begin{equation}
t=-\frac{M_{2}^{3}}{g^{2}}\,\,\int^{z}dz^{\prime }\left( z^{\prime }\ln
\left| \frac{z^{\prime }}{ez_{0}}\right| -z_{1}\ln \left| \frac{z_{1}}{ez_{0}%
}\right| \right) ^{-2}
\end{equation}
determining the dependence of $z$ from $t$.

Because the integral over $z^{\prime }$ is divergent at $z^{\prime }=z_{1}$,
we conclude, that at $t\rightarrow \pm \infty $ the coordinates $l$ and $%
r_{1}$ grow linearly with $t$, but their ration $z$ tends to the constants $%
z_{1}$ or $\widetilde{z}_{1}$, where $\widetilde{z}_{1}$ is the second
solution of the equation
\begin{equation}   
\widetilde{z}_{1}\ln \left| \frac{\widetilde{z}_{1}}{ez_{0}}\right|
=z_{1}\ln \left| \frac{z_{1}}{ez_{0}}\right|
\end{equation}
appearing if $|z_{1}|<|ez_{0}|$. Really the particle does not perform any
oscillation before going to infinity. We show below, that this instability
takes place in the general case.

Using the above expression for the angular momenta one can express $l$ in
terms of the coordinates $r_{1}$ and $r_{2}$
\begin{equation}
l=-\frac{M_{1}}{M_{0}}\,r_{1}-\frac{M_{2}}{M_{0}}\,r_{2}\,.
\end{equation}
Because $M_{0}$ is conserved, it is enough to consider only the first vector
component of the Newton equation
\begin{equation}
\partial _{t}^{2}\,r_{1}=\frac{g}{r_{1}^{2}}\,\frac{1}{(1+y^{2})\,\left(
\frac{M_{1}}{M_{0}}\,+\frac{M_{2}}{M_{0}}\,y\right) }\,\,,
\end{equation}     
where we introduced the new variable $y$%
\begin{equation}
y=\frac{r_{2}}{r_{1}}\,.
\end{equation}

In the variables $r_{1}$ and $y$ the integral of motion $M_{0}$ has the form

\begin{equation}
M_{0}=r_{1}^{2}\,\partial _{t}\,y
\end{equation}  
and therefore the equation for $r_{1}$ can be written as follows

\begin{equation}
\partial _{t}^{2}\,r_{1}=\frac{g}{M_{0}}\,\,\frac{\partial _{t}\,y}{%
(1+y^{2})\left( \frac{M_{1}}{M_{0}}\,+\frac{M_{2}}{M_{0}}\,y\right) }\,.
\end{equation}
By integrating it we obtain the fourth integral of motion 
\begin{equation}   
\partial _{t}\,r_{1}=\frac{g}{M_{0}}\,\int_{y_{0}}^{y}\frac{d\,\widetilde{y}%
}{(1+\,\widetilde{y}^{2})\left( \frac{M_{1}}{M_{0}}\,+\frac{M_{2}}{M_{0}}\,%
\widetilde{y}\right) }\,,
\end{equation}
where the constant $y_{0}$ is equal to the value of $r_{2}/r_{1}$ for which $%
\partial _{t}\,r_{1}=0$.

Now it is convenient to use again the integral of motion $M_{0}$ to express $%
r_{1}$ in terms of $y$

\begin{equation}
-\frac{1}{2\,\partial _{t}\,y}\,\sqrt{\frac{M_{0}}{\partial _{t}\,y}}%
\,\partial _{t}^{2}\,y=\frac{g}{M_{0}}\,\int_{y_{0}}^{y}\frac{d\,\widetilde{y%
}}{(1+\,\widetilde{y}^{2})\left( \frac{M_{1}}{M_{0}}\,+\frac{M_{2}}{M_{0}}\,%
\widetilde{y}\right) }\,.
\end{equation}
This operation allows one to find the fifth integral of motion
\begin{equation}
\sqrt{M_{0}\,\partial _{t}\,y\,}=-\frac{g}{M_{0}}\,\int_{y_{1}}^{y}d\,z%
\int_{y_{0}}^{z}\frac{d\,\widetilde{y}}{(1+\,\widetilde{y}^{2})\left( \frac{%
M_{1}}{M_{0}}\,+\frac{M_{2}}{M_{0}}\,\widetilde{y}\right) }\,,
\end{equation}
where the constant $y_{1}$ is the value of the ration $r_{2}/r_{1}$ for
which $\partial _{t}\,y=0$. We can write this integral of motion in the form

\begin{equation}
\frac{M_{0}}{r_{1}}=\sqrt{M_{0}\,\partial _{t}\,y\,}=-\frac{g}{M_{0}}%
\,\int_{y_{0}}^{y}\frac{d\,\widetilde{y}\,\,(y-\max \,(y_{1},\widetilde{y}))%
}{(1+\,\widetilde{y}^{2})\left( \frac{M_{1}}{M_{0}}\,+\frac{M_{2}}{M_{0}}\,%
\widetilde{y}\right) }\,.
\end{equation}
The last equation gives a possibility to calculate the particle trajectory   
parametrized by $y$

\begin{equation}
r_{1}=\frac{r_{2}}{y}=-\frac{l}{\frac{M_{1}}{M_{0}}\,+\frac{M_{2}}{M_{0}}%
\,y\,}=-\frac{|M_{0}|}{g}\left( \int_{y_{0}}^{y}\frac{d\,\widetilde{y}%
\,\,(y-\max \,(y_{1},\widetilde{y}))}{(1+\,\widetilde{y}^{2})\left(
M_{1}\,+M_{2}\,\widetilde{y}\right) }\right) ^{-1}.
\end{equation}

To find the coordinate dependence from time one should invert the equation  

\begin{equation}
t=\frac{M_{0}}{g^{2}}\int^{y(t)}dz\,\,\left( \int_{y_{0}}^{z}\frac{d\,%
\widetilde{y}\,\,(z-\max \,(y_{1},\widetilde{y}))}{(1+\,\widetilde{y}%
^{2})\left( M_{1}\,+M_{2}\,\widetilde{y}\right) }\right) ^{-2}.
\end{equation}
Thus, in the general case we see, that due to the divergency of the integral
over $z$ at $z\rightarrow y_{1}$ the particle in the ternary potential at $%
t\rightarrow \pm \infty $ goes to infinity along the line with a fixed ratio
of coordinates
\begin{equation}
r_{1}=\frac{r_{2}}{y_{1}}=-\frac{l}{\frac{M_{1}}{M_{0}}\,+\frac{M_{2}}{M_{0}}%
\,y_{1}\,}\sim t.
\end{equation}
The reason for this instability is the reflective character of the
ternary force at $l<0$. In
principle a particle or a large cosmic object with the magnetic field
of the ternary type can exist in nature. Therefore it is interesting to     
calculate the differential cross-section for the particle scattering off
such field.

\subsection{Monopole scattering in the ternary field}

Here we investigate the scattering of the monopole in the magnetic field $h$
introduced above. Initially we consider the symmetric case, where the
trisectrice lies in the scattering plane ($r_{1},l$) fixed by the angular  
momentum $\overrightarrow{M}=(0,M_{2},0)$. The monopole trajectory
approaches the critical slope $z=l/r_{1}\rightarrow z_{1}$ only
asymptotically  
\begin{equation}
r_{1}\rightarrow \frac{M_{2}^{2}}{g}\,\frac{1}{(z-z_{1})\,\ln \left| \frac{%
z_{1}}{z_{0}}\right| }\,,\,\,\partial _{t}\,r_{1}\rightarrow \frac{g}{M_{2}}%
\,\,\ln \left| \frac{z_{1}}{z_{0}}\right| \,,\,\,\frac{l}{r_{1}}\rightarrow
z_{1}\,.
\end{equation}
Apart from $M_2$ one can fix for the colliding monopole also the
asymptotic slope $z_{1}$ and its velocity
\begin{equation}
v_{1}^{-\infty }=\lim_{t\rightarrow -\infty }\partial _{t}\,r_{1}\,.
\end{equation}
The velocity component along the axis $r_{0}=l$ is calculated in terms of $%
v_{1}^{-\infty }$ and $z_{1}$%
\begin{equation}
v_{0}^{-\infty }=v_{1}^{-\infty }z_{1}\,.
\end{equation}
Together with $M_{2}$ the total velocity determines the impact parameter $%
b=M_{2}/|v^{-\infty }|$. Note, that the velocity vector at $t\rightarrow
-\infty $ should be directed to the center
\begin{equation}
\frac{\overrightarrow{v}}{|v|}=-\frac{\overrightarrow{R}}{|R|}\,.
\end{equation}
The initial conditions fix also the parameter $z_{0}$
\begin{equation}
v_{1}^{-\infty }=\,\frac{g}{M_{2}}\,\,\ln \left| \frac{z_{1}}{z_{0}}\right| .
\end{equation}
Due to the invariance of the equations under the time inversion $t\rightarrow -t
$ and the reflection $r_{1}\rightarrow -r_{1}$ it is enough to
investigate only the positive values of $M_{2}$ and $z_{1}$

\begin{equation}
M_{2}>0\,,\,\,z_{1}>0\,.
\end{equation}

Let us consider initially the case $z_{0}<z_{1}$.

The asymptotic slope $z=z_{1}$ can be reached at $t=-\infty $ or $t=\infty $%
. \ At the first sub-case the monopole at large negative $t$ moves to the
center along the line $l/r_{1}=z_{1}-0$ from the negative values of $l$ and $%
r_{1}$ with the initial velocity ($v_{1}^{-\infty },v_{0}^{-\infty })$. The
velocity $v_{1}$ changes its sign at $z=z_{0}$ \ Providing that $z_{1}<ez_{0}
$ corresponding to the condition $v_{1}^{-\infty }<g/M_{2}$, the monopole
turns before reaching the axes $r_{1}=0$, $l=0$ and moves backward
along the asymptotic line $z=\widetilde{z}_{1}<z_{0}$ being another
solution of the equation
\begin{equation}
\widetilde{z}_{1}\ln \left| \frac{\widetilde{z}_{1}}{ez_{0}}\right|
=z_{1}\ln \left| \frac{z_{1}}{ez_{0}}\right| \,.
\end{equation}  
In particular, we have $\widetilde{z}_{1}\rightarrow z_{0}-\epsilon $ for $%
z_{1}\rightarrow z_{0}+\epsilon $ and $\widetilde{z}_{1}\rightarrow \epsilon
ez_{0}/\ln (1/\epsilon )$ for $z_{1}\rightarrow ez_{0}-\epsilon $. Note,
that for $M_{2}<0$ the monopole goes along the same trajectory but in the
opposite direction if one would interchange $z_{1}\leftrightarrow \widetilde{%
z}_{1}$.

If $z_{1}>ez_{0}$ the particle approaches at $z=0$ the line $l=0$ where its   
velocity $v_{1}$ tends to $-\infty $. In this moment we have

\begin{equation}
\lim_{l\rightarrow 0}r_{1}=-\frac{M_{2}^{2}}{g}\,\frac{1}{z_{1}\ln \left|
\frac{z_{1}}{ez_{0}}\right| }<0\,.
\end{equation}
After that the velocity $|v_{1}|$ decreases and vanishes at $z=-z_{0}$. At $%
z\rightarrow -\infty $ the monopole trajectory approaches the point $%
r_{1}=-0\,,\,l=+0$
\begin{equation}
\lim_{z\rightarrow -\infty }r_{1}\rightarrow \frac{M_{2}^{2}}{g}\,\frac{1}{%
z\ln \left| \frac{z}{ez_{0}}\right| }\,,\,\lim_{z\rightarrow -\infty
}\,l\rightarrow \frac{M_{2}^{2}}{g}\,\frac{1}{\ln \left| \frac{z}{ez_{0}}%
\right| }
\end{equation}
Simultaneously the velocities grow rapidly
\begin{equation}
\lim_{z\rightarrow -\infty }v_{1}=\frac{g}{M_{2}}\,\ln \left| \frac{z}{z_{0}}%
\right| \,,\,\,\lim_{z\rightarrow -\infty }v_{0}=\frac{g}{M_{2}}\,z\,\ln
\left| \frac{z}{z_{0}}\right|
\end{equation}  
and the monopole reaches the coordinate center with an infinite energy for a
finite period of time. One can prolong the trajectory in the sector $%
l<0,\,r_{1}>0$ where after the change of the velocity sign $v_{1}/|v_{1}|$
at $z=-z_{0}$ the particle goes asymptotically to $-\infty $ along the axes $%
l$. In the second sub-case the monopole moves to $r_{1}=l=\infty $ along the
line $z=z_{1}+0$ from the coordinate center $r_{1}=+0\,,\,l=+0,\,l/r_{1}=%
\infty $.

In the case $z_{0}<z_{1}$ we have $v_{1}<0$ for $z\rightarrow z_{1}$ and
therefore the monopole can arrive from $r_{1}=l/z_{1}=\infty $ or depart to $%
r_{1}=l/z_{1}=-\infty $. In the first sub-case it reaches the line $l=0$ at $%
z=0$ with the velocity $v_{1}=-\infty $ in the point
\begin{equation}
\lim_{l\rightarrow 0}r_{1}=-\frac{M_{2}^{2}}{g}\,\frac{1}{z_{1}\ln \left|
\frac{z_{1}}{ez_{0}}\right| }\,.
\end{equation}
Then in the point $z=-z_{0}$ its velocity $v_{1}$ changes its sign and the
monopole goes to $r_{1}=\infty $ at $t\rightarrow \infty $ for the value $%
z^{\prime }<0$ which is a solution of the equation

\begin{equation}
z^{\prime }\ln \left| \frac{z^{\prime }}{ez_{0}}\right| =z_{1}\ln \left|
\frac{z_{1}}{ez_{0}}\right| \,.
\end{equation}
In the second sub-case the monopole arises at $r_{1}=-0,\,l=-\infty $ at $%
z=-\infty $ and goes to $-\infty $ at $z\rightarrow z_{1}-0$.

Let us consider now the movement of the monopole in the general case, when
the scattering plane does not contain the trisectrice. The initial data
include the three components of angular momentum $\overrightarrow{M}$, and
velocity components
\begin{equation}
v_{1}^{-\infty }=\frac{v_{2}^{-\infty }}{y_{1}}=-\frac{v_{0}^{-\infty }}{%
\frac{M_{1}}{M_{0}}+\frac{M_{2}}{M_{0}}y_{1}}\,,
\end{equation}
where $y_{1}$ is the integral of motion. It means, that the angular momenta
$M_i$ satisfy the relation
\begin{equation}
M_1v_1^{-\infty }+M_2v_2^{-\infty }+M_0v_0^{-\infty }=0
\end{equation}
and only two of them ($M_1,\,M_2$) are independent.

Note, that due to the symmetry
to
the rotations around the axes $l$ without any restriction of generality one
can put
\[
v_{2}^{-\infty }=y_{1}=0\,.
\]
In this case the  introduced above parameter $z_{1}$ is
\[
z_{1}=-\lim_{y_{1},M_{0}\rightarrow 0}\left( \frac{M_{1}}{M_{0}}+\frac{M_{2}%
}{M_{0}}y_{1}\right) \,.
\]

The other integral of motion $y_{0}$ is fixed by the equation
\[
v_{1}^{-\infty }=\frac{g}{M_{0}}\,\int_{y_{0}}^{y_{1}}\frac{d\,\widetilde{y}%
}{(1+\,\widetilde{y}^{2})\left( \frac{M_{1}}{M_{0}}\,+\frac{M_{2}}{M_{0}}\,%
\widetilde{y}\right) }
\]
  
\begin{equation}
=g\left( \frac{i/2}{M_{1}\,-\,iM_{2}}\,\ln \frac{y_{1}-i}{y_{0}-i}-\frac{i/2%
}{M_{1}\,+\,iM_{2}}\,\ln \frac{y_{1}+i}{y_{0}+i}+\frac{M_{2}}{%
M_{1}^{2}+M_{2}^{2}}\,\ln \frac{M_{1}\,+y_{1}M_{2}}{M_{1}\,+y_{0}M_{2}}%
\right) .
\end{equation}
It allows one to calculate the velocity $v_{1}$ as function of the parameter
$y=r_{2}/r_{1}$%
\[
v_{1}(y)
\]
\begin{equation}
=g\left( \frac{i/2}{M_{1}\,-\,iM_{2}}\,\ln \frac{y-i}{y_{0}-i}-\frac{%
i/2}{M_{1}\,+\,iM_{2}}\,\ln \frac{y+i}{y_{0}+i}+\frac{M_{2}}{%
M_{1}^{2}+M_{2}^{2}}\,\ln \frac{M_{1}\,+yM_{2}}{M_{1}\,+y_{0}M_{2}}\right)  
\end{equation}
and the trajectory
\begin{equation}
r_{1}=\frac{r_{2}}{y}=-\frac{l}{\frac{M_{1}}{M_{0}}\,+\frac{M_{2}}{M_{0}}%
\,y\,}=-\frac{M_{0}}{g}\left( \int_{y_{1}}^{y}dy^{\prime }\frac{%
v_{1}(y^{\prime })}{g}\right) ^{-1}.
\end{equation}
Here
\[
\int^{y}dy^{\prime }\frac{v_{1}(y^{\prime })}{g}
\]
\[
=\frac{i(y-i)/2}{%
M_{1}\,-\,iM_{2}}\,\ln \frac{y-i}{e(y_{0}-i)}-\frac{(y+i)i/2}{%
M_{1}\,+\,iM_{2}}\,\ln \frac{y+i}{e(y_{0}+i)}+\frac{M_{1}\,+yM_{2}}{%
M_{1}^{2}+M_{2}^{2}}\,\ln \frac{M_{1}\,+yM_{2}}{e(M_{1}\,+y_{0}M_{2})}\,\,.
\]
The above formulas give a possibility to calculate the differential
cross-section for the scattering of the monopole moving at $t\rightarrow
-\infty $ with the velocity $\overrightarrow{v}^{-\infty }$ to the
coordinate center
\begin{equation}
d\sigma =d^{2}\rho =\rho d\rho d\varphi \,,
\end{equation}
where $\overrightarrow{\rho }$ is impact parameter and $\varphi $ is its
azimuthal angle around the velocity $\overrightarrow{v}^{-\infty }$. We have
the following relation between $\overrightarrow{M}$ and
$\overrightarrow{\rho }$
\begin{equation}
\overrightarrow{M}=\left[ \overrightarrow{\rho },\overrightarrow{v}^{-\infty
}\right] \,,\,\,\overrightarrow{\rho }=\frac{1}{|\overrightarrow{v}^{-\infty
}|^{2}}\left[ \overrightarrow{v}^{-\infty },\overrightarrow{M}\right] \,,
\end{equation}
where we took into account, that $(\overrightarrow{\rho },\overrightarrow{v}%
^{-\infty })=0$. As it was argued above, one can put $y=v_2^{-\infty}=0$ without
the generality loss. In this case the above relations are simplified
\begin{equation}
\rho ^{pl} |\vec{v}^{-\infty}|=M_2 \,,\,\,\rho ^\perp |\vec{v}^{-\infty}|^2=
M_1v_0^{-\infty}-M_0v_1^{-\infty}
=M_1\,
\frac{|\vec{v}^{-\infty}|^2}{v_0^{-\infty}} \,,
\end{equation}
where we introduced the components $\rho ^{pl}$ and $\rho ^\perp$ belonging
to the plane ($r_1,l$) and orthogonal to it, respectively. These expressions
give a possibility to calculate the Jacobian of the transition
from the vector $\vec{\rho}$ to the variables $M_{1,2}$
\begin{equation}
d^2\rho = \frac{dM_1\,dM_2}{|v_0^{-\infty}|
|\vec{v}^{-\infty}|}\,.
\end{equation}

Further, the parameters of the final states can be expressed in terms of
angular momenta $M_1,M_2$. For example, if the monopole moves
from $r_{1}=r_{2}=l=-\infty $ with such small velocity, that the equation
\begin{equation}
\int_{y_{1}}^{y}dy^{\prime }\frac{v_{1}(y^{\prime })}{g}=0
\end{equation}
has the other solution $y=\widetilde{y}_{1}(M_1,M_2)$
apart from trivial one $y=y_{1}$, than the momenta of the particle at
$t\rightarrow \infty$ are related by the formulas
\begin{equation}
v_{1}^{\infty }=\frac{v_{2}^{\infty }}{\widetilde{y}_{1}}=
-\frac{v_{0}^{\infty }}{%
\frac{M_{1}}{M_{0}}+\frac{M_{2}}{M_{0}}\widetilde{y}_{1}}\,,
\end{equation}
The energy is not conserved in the ternary field and therefore the
final kinetic energy $E$ is also a function of the angular momenta $M_1,M_2$
\begin{equation}
E(M_1,M_2)=\frac{(\vec{v}^{\infty})^2}{2}
\end{equation}
and can be obtained from the previous formulas.
Therefore we can calculate the Jacobian of the transition from the
momenta ($M_1,M_2$) to the variables ($\widetilde{y}_{1},E$)
\begin{equation}
d\widetilde{y}_{1}\,dE=J\,dM_1\,dM_2\,,\,\,J=\det \left(
\begin{array}{cc}
\frac{\partial \widetilde{y}_{1}}{\partial \,M_{1}} & \frac{\partial 
\widetilde{y}_{1}}{\partial \,M_2} \\
\frac{\partial E}{\partial \,M_1} & \frac{\partial
E}{\partial \,M_2}
\end{array}
\right) \,.
\end{equation}
Thus, the differential cross-section for small momenta of the monopole
moving to the center from negative values of $r_1,\,r_2,\,l$ is
\begin{equation}
d\sigma =\frac{d\widetilde{y}_{1}\,dE}{J\,
|\vec{v}_0^{-\infty}|\,|\vec{v}^{-\infty}|}\,.
\end{equation}

{\bf Acknowledgments} One of us (GGV) acknowledges Laboratoire de Physique 
Th\'eorique, Strasbourg and LAPTH Annecy-Le-Vieux, where part of the work 
was done, for hospitality. He also thanks for discussions and
support L.Alvarez-Gaume, U. Aglietti, I. Antoniadis,  G. Belanger, G. Costa, 
I. Crotty, B. Dubrovin, A. Dubrovskiy, J.Ellis, T.Faberge, 
L.Fellin, G. Girardi, 
M. Goze, R. Kerner, A. Masiero, 
J. Richert, 
A. Sabio-Vera,
P. Sorba, A. Uranga and J.-B. Zuber.


\begin{thebibliography}{9}
\bibitem{appell} P. Appell, CR, Acad. Sci., Paris {\bf 84} (1877) 540,
P. Appell, CR, Acad. Sci., Paris {\bf 84} (1877) 1378.
%
\bibitem{humbert1}
P. Humbert, J. Math Pures Appl. {\bf 21}  (1942) 141.
%
\bibitem{humbert2}
P. Humbert, Bull. Math {\bf 66} (1942) 145.
%
\bibitem{humbert3}
P. Humbert, Bull. Math {\bf 68} (1944) 50.
%
%
\bibitem{devisme1} J. Devisme, Annales Fac. Sci. Toulouse { \bf 25}
 (1933) 143.
%
\bibitem{devisme2} J. Devisme, J. Math. Pures Appl. {\bf 19}
(1940) 359.
%
\bibitem{mc1}  N.~Fleury, M.~Rausch de Traubenberg and R.~M.~Yamaleev, 
J.\ Math.\ Anal.\ Appl.\ \textbf{180} (1993) 431. 
%

\bibitem{mc2}  N.~Fleury, M.~Rausch de Traubenberg and R.~M.~Yamaleev, 
J.\ Math.\ Anal.\ Appl.\ \textbf{191} (1995) 118. 
%
\bibitem{kerner}
R.~Kerner, Class. Quantum Grav. \textbf{14},  (1997) A203.
%
\bibitem{volkov}
G. G. Volkov, 
{Annales Fond.  Broglie} {\bf 31}  (2006) 227
[arXiv:hep-ph/0607334].
%
\bibitem{jones}   H.~F. Jones,
{Groups, representations and physics},   Bristol, UK,  Hilger (1990).
\bibitem{int}  P.~Baseilhac, S.~Galice, P.~Grange and M.~Rausch de
Traubenberg, 
Phys.\ Lett.\ B \textbf{478} (2000) 365 [arXiv:hep-th/0002232]; 
P.~Baseilhac, P.~Grange and M.~Rausch de Traubenberg, 
Mod.\ Phys.\ Lett.\ A \textbf{13}  (1998) 2531. 
%

\bibitem{forster}  Otto Forster, {Lectures on Riemann Surfaces} Springer
(1981).
%
\bibitem{mf}  A. Michenko and Fomenko, {A course of differential geometry
and topology}, Mir Publishers Moscow, (1988).
%

\end{thebibliography}
\end{document}